\documentclass[preprint]{aastex}

\newcommand{\half}{\frac{1}{2}}

\received{15 November 2000}
\accepted{\underline{\hspace{50mm}}}
\slugcomment{Accepted by \apjs\ / 2001 Feb 07}
\shortauthors{Turner \& Stone}
\shorttitle{Radiation Hydrodynamics with ZEUS-2D}

\begin{document}
\title{A Module for Radiation Hydrodynamic Calculations With ZEUS-2D
Using Flux-Limited Diffusion}

\author{N. J. Turner\altaffilmark{1} \& J. M. Stone\altaffilmark{1}}

\altaffiltext{1}{Astronomy Department, University of Maryland, College
Park MD 20742, U.\ S.\ A.; {\tt neal@astro.umd.edu}}

\begin{abstract}
A module for the ZEUS-2D code is described which may be used to solve
the equations of radiation hydrodynamics to order unity in $v/c$, in
the flux-limited diffusion (FLD) approximation.  In this
approximation, the tensor Eddington factor $\mathsf f$ which closes
the radiation moment equations is chosen to be an empirical function
of the radiation energy density.  This is easier to implement and
faster than full-transport techniques, in which $\mathsf f$ is
computed by solving the transfer equation.  However, FLD is less
accurate when the flux has a component perpendicular to the gradient
in radiation energy density, and in optically thin regions when the
radiation field depends strongly on angle.

The material component of the fluid is here assumed to be in local
thermodynamic equilibrium.  The energy equations are operator-split,
with transport terms, radiation diffusion term, and other source terms
evolved separately.  Transport terms are applied using the same
consistent transport algorithm as in ZEUS-2D.  The radiation diffusion
term is updated using an alternating-direction implicit method with
convergence checking.  Remaining source terms are advanced together
implicitly using numerical root-finding.  However when absorption
opacity is zero, accuracy is improved by instead treating the
compression and expansion source terms using a time-centered
differencing scheme.

Results are discussed for test problems including radiation-damped
linear waves, radiation fronts propagating in optically-thin media,
subcritical and supercritical radiating shocks, and an optically-thick
shock in which radiation dominates downstream pressure.
\end{abstract}

\keywords{hydrodynamics --- methods: numerical --- radiative transfer}

\section{INTRODUCTION}

In many astrophysical systems, radiation is the dominant energy
transport mechanism.  In some circumstances, including winds from
Wolf-Rayet stars (Lucy \& Abbott 1993) and asymptotic giant branch
stars (Habing 1996), interiors of massive stars (Kippenhahn \& Weigert
1990), supernova blast waves (Arnett et al. 1989), and compact object
accretion flows (Klein et al. 1996) and disks (Shakura \& Sunyaev
1973), radiation comprises a substantial fraction of the total energy
density, momentum density, or pressure.  The dynamics of such flows
must be modeled using a radiation hydrodynamical (RHD) approach, in
which energy and momentum conservation laws for the radiation field
are solved along with those for the gas.  The outcomes of these
calculations depend on the exchange of internal energy and momentum
between material and radiation.  The length and time scales associated
with the exchanges are determined by atomic processes, and may be
orders of magnitude shorter or faster than dynamical scales, so that
implicit time differencing is usually necessary.  An accurate
description of the angular dependence of the radiation field may be
important when parts of the gas are optically thin, localized sources
of radiation are present, or shadows are cast within the flow.  It is
possible to compute the angular dependence by solving the transfer
equation along rays through the points of interest.  However, the
number of rays needed to adequately specify the angular dependence at
every point is often large.  Such a method has been implemented in two
spatial dimensions in the ZEUS-2D magnetohydrodynamics (MHD) code
(Stone \& Norman 1992b) by Stone, Mihalas, \& Norman (1992), but is
complex and computationally intensive.  Despite the promise of this
full-transport method, it must be considered still under development.
In this paper we describe a method for calculations using ZEUS-2D, in
which the angular dependence of the radiation field is assumed to be
given by the flux-limited diffusion (FLD) approximation.  This method
is simple, robust, and relatively cheap in computer time.  We assume
the gas is in local thermodynamic equilibrium (LTE) at a temperature
which need not correspond to that of the radiation field.  Frequency
dependence of the opacities is neglected, though it may be included in
a straightforward fashion.  The RHD equations solved are discussed in
\S\ref{sec:eqns}, the FLD approximation in \S\ref{sec:fld}, and the
numerical algorithm in \S\ref{sec:method}.  Results of test
calculations are presented in \S\ref{sec:tests}, and the advantages
and limitations of the method are summarized in \S\ref{sec:summary}.

\section{EQUATIONS OF RADIATION HYDRODYNAMICS\label{sec:eqns}}

Much as the gas dynamical equations are derived by taking velocity
moments of the material particle kinetic equation, or Boltzmann
equation, the conservation laws for the radiation field are generated
by taking angular moments of the photon kinetic equation, or radiation
transport equation.  In a frame comoving with the radiating fluid,
assuming LTE, and to order unity in $v/c$, the coupled equations of
RHD are (Mihalas \& Mihalas 1984)
\begin{equation}\label{eqn:cty}
{D\rho\over D t}+\rho\nabla\cdot{\bf v}=0,
\end{equation}
\begin{equation}\label{eqn:gasmomentum}
\rho{D{\bf v}\over D t} = -\nabla p + {1\over c}\chi_F{\bf F},
\end{equation}
\begin{equation}\label{eqn:radenergy}
\rho{D\over D t}\left({E\over\rho}\right) =
	- \nabla\cdot{\bf F} - \nabla{\bf v}:{\mathsf P}
	+ 4\pi\kappa_P B - c\kappa_E E,
\end{equation}
\begin{equation}\label{eqn:gasenergy}
\rho{D\over D t}\left({e\over\rho}\right) =
	- p \nabla\cdot{\bf v}
	- 4\pi\kappa_P B + c\kappa_E E,
\end{equation}
and
\begin{equation}\label{eqn:radmomentum}
{\rho\over c^2}{D\over D t}\left({{\bf F}\over\rho}\right) =
	- \nabla\cdot{\mathsf P}
	- {1\over c}\chi_F{\bf F}.
\end{equation}
Here the convective derivative $D/Dt\equiv\partial/\partial t + {\bf
v}\cdot\nabla$.  The dependent quantities $\rho$, $e$, ${\bf v}$, and
$p$ are the material mass density, energy density, velocity, and
scalar isotropic pressure respectively, while $E$, ${\bf F}$, and
${\mathsf P}$ are the total frequency-integrated radiation energy
density, momentum density or flux, and pressure tensor, respectively.
The latter are the zeroth, first, and second angular moments of the
radiation specific intensity.  The specific intensity, $I({\bf x},
{\bf\Omega}, \nu, t)$, in general varies with spatial position ${\bf
x}$, viewing direction ${\bf\Omega}$, frequency $\nu$, and time $t$.
Complete description of $I$ in a three-dimensional space therefore
requires seven scalar coordinates.  The angular moments of $I$
appearing in the RHD equations above may be written
\begin{equation}\label{eqn:emoment}
E({\bf x}, t)={1\over c}\int_{0}^{\infty}d\nu
	\oint d\Omega\, I({\bf x}, {\bf \Omega}, \nu, t),
\end{equation}
\begin{equation}\label{eqn:fmoment}
{\bf F}({\bf x}, t)=\int_{0}^{\infty}d\nu
	\oint d\Omega\, I({\bf x}, {\bf \Omega}, \nu, t) {\bf n},
\end{equation}
and
\begin{equation}\label{eqn:pmoment}
{\mathsf P}({\bf x}, t)={1\over c}\int_{0}^{\infty}d\nu
	\oint d\Omega\, I({\bf x}, {\bf \Omega}, \nu, t) {\bf n n}.
\end{equation}

The assumption of LTE allows the source function specifying the rate
of emission of radiation from the gas in
equations~(\ref{eqn:radenergy}) and~(\ref{eqn:gasenergy}) to be
written as the Planck function $B$.  Equations~(\ref{eqn:gasmomentum})
through~(\ref{eqn:radmomentum}) have been integrated over frequency,
leading to the flux mean total opacity $\chi_F$, and the Planck mean
and energy mean absorption opacities, $\kappa_P$ and $\kappa_E$.  In
the present paper the opacities are assumed to be independent of
frequency, so that $\kappa_P=\kappa_E$ and the subscripts may be
omitted.  The total opacity $\chi$ is the sum of components due to
absorption, $\kappa$, and scattering, $\sigma$, all having dimensions
of inverse length.  In a pure hydrogen gas above $10^4$~K, simple
forms for $\kappa$ and $\sigma$ are Kramers' Law and the Thomson
scattering coefficient.  Equations~(\ref{eqn:cty})
to~(\ref{eqn:radmomentum}) correctly describe the flow only in the
comoving frame, where material properties are isotropic and the form
of the material-radiation interaction terms is greatly simplified.  In
writing the equations, effects of self-gravity and magnetic fields
have been ignored.  The treatment of these effects in ZEUS-2D is
independent of the algorithms for RHD described here.

The equations of RHD may be closed by the addition of constitutive
relations for the Planck function and opacities, an equation of state
specifying the gas pressure, and an assumption about the relationship
between the angular moments of the radiation field.  Since ${\mathsf
P}=\frac{1}{3}E$ when the field is isotropic, one might choose to
assume this relation holds everywhere.  This is referred to as the
Eddington approximation, and is analogous to assuming isotropy of the
material particle distribution function in deriving the gas equation
of state.  The Eddington approximation implies that in steady-state,
equation~(\ref{eqn:radmomentum}) becomes
\begin{equation}\label{eqn:eddflux}
{\bf F}=-{c\over 3\chi}\nabla E.
\end{equation}
This expression gives the correct flux in optically thick regions,
where $\chi$ is large, and tends to infinity in optically thin regions
where $\chi\rightarrow 0$.  However the rate at which radiation
transports energy is finite and, from equations~(\ref{eqn:emoment})
and~(\ref{eqn:fmoment}), is limited to $|{\bf F}|\leq c E$.  This
causality problem with equation~(\ref{eqn:eddflux}) occurs because in
optically thin regions photons travel freely, their mean free paths
may exceed characteristic lengths in the system, and the radiation
field may be anisotropic.  In this paper we implement an extension of
the Eddington approximation in which causality is preserved by
assuming a particular form for the angular dependence of the radiation
field in optically-thin regions.  This extension is described in the
next section.

\section{THE FLUX-LIMITED DIFFUSION APPROXIMATION\label{sec:fld}}

FLD techniques were first used in astrophysics to solve the radiation
transport equation, in calculations of accretion onto a neutron star
(Alme \& Wilson 1974).  Subsequently, Levermore \& Pomraning (1981;
hereafter LP) generalized the method to approximately handle transport
phenomena while preserving causality in regions where spatial
variations occur over distances smaller than a mean free path.
Pomraning (1983) included relativistic terms of order $v/c$ in a
moving fluid.  FLD methods have since been used in pure scattering
media (Melia \& Zylstra 1991), and for fully general-relativistic
calculations (Anile \& Romano 1992).  The key ingredient in all cases
is the assumption that the specific intensity is a slowly-varying
function of space and time.  In one space dimension, this assumption
is valid in both the optically-thick diffusion limit and the
optically-thin free-streaming limit.  One hopes it holds approximately
in the intermediate regime and in multi-dimensions, and this may be
tested by comparison with results of full-transport calculations.
Given slow spatial and time variation, analytic relations may be
obtained between the angular moments of the radiation field.  LP
showed that the radiation flux may be written in the form of Fick's
law of diffusion, as
\begin{equation}\label{eqn:fick}
{\bf F}=-D\nabla E,
\end{equation}
with a diffusion coefficient $D$ given by
\begin{equation}\label{eqn:dc}
D={c\lambda\over\chi}.
\end{equation}
The dimensionless function $\lambda=\lambda(E)$ is called the flux
limiter.  In this framework, the radiation pressure tensor may be
expressed in terms of the radiation energy density via
\begin{equation}\label{eqn:eddtensor}
{\mathsf P}={\mathsf f}E,
\end{equation}
where the components of the Eddington tensor $\mathsf f$ are given
by 
\begin{equation}\label{eqn:eddfactor}
{\mathsf f}=\frac{1}{2}(1-f){\mathsf I} + \frac{1}{2}(3f-1){\bf n n},
\end{equation}
${\bf n}=\nabla E/|\nabla E|$ is the unit vector in the direction of
the radiation energy density gradient, and the dimensionless scalar
function $f=f(E)$ is called the Eddington factor.  The flux limiter
$\lambda$ and Eddington factor $f$ are related through implicit
constraints between the moments ${\bf F}$ and ${\mathsf P}$, so that
\begin{equation}\label{eqn:flambda}
f=\lambda+\lambda^2 R^2,
\end{equation}
where $R$ is the dimensionless quantity $R=|\nabla E|/(\chi E)$.

\begin{figure}
\plotone{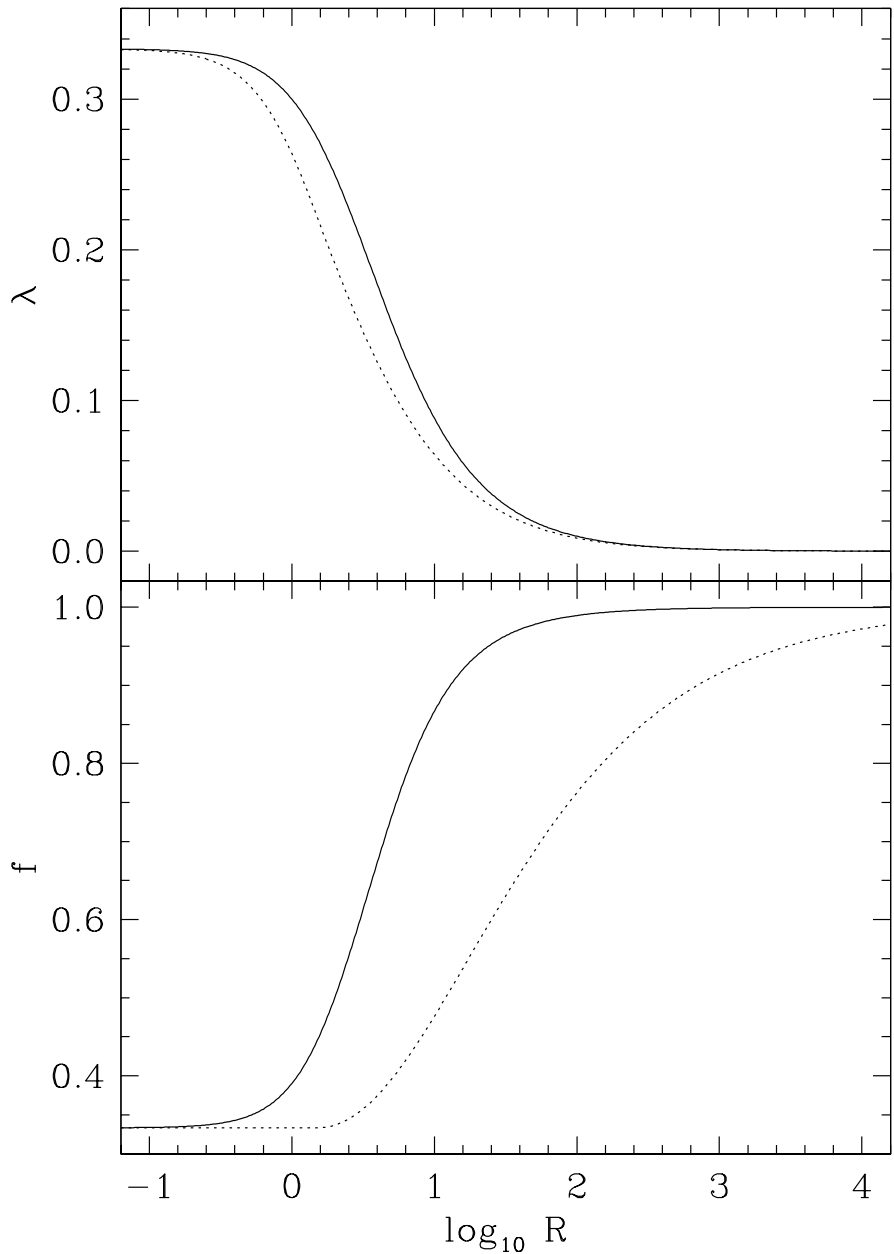}
\caption{\label{fig:limiters}Dependence of the flux-limiter (upper
panel) and Eddington factor (lower panel) on optical thickness
parameter $R=|\nabla E|/(\chi E)$.  Levermore-Pomraning values
(eq.~[\ref{eqn:lplimiter}]) are shown by solid curves, Minerbo
(eq.~[\ref{eqn:minerbolimiter}]) by dotted curves.  Eddington factors
are computed from the flux-limiters using
equation~(\ref{eqn:flambda}).  }
\end{figure}

Equations~(\ref{eqn:fick}) through~(\ref{eqn:flambda}) close the
equations of RHD, eliminating the need to solve the radiation momentum
equation~(\ref{eqn:radmomentum}), and greatly simplifying the
computation.  Since the angular distribution of the radiation field is
not explicitly computed, no formal solution of the transfer equation
is needed.  An important choice which must be made to achieve this
simplification is the expression to be used for the flux limiter
$\lambda$.  Many expressions are possible which preserve causality,
and are consistent with the assumption of smoothness in the radiation
field.  These are distinguished by different assumptions regarding the
details of the angular dependence of the specific intensity.  In one
example, LP chose to constrain the normalized specific intensity
$\psi=I/(c E)$ by
\begin{equation}\label{eqn:lpangular}
{1\over c}{\partial\psi\over\partial t} + {\bf\Omega}\cdot{\nabla\psi}
= 0,
\end{equation}
resulting in the flux limiter
\begin{equation}\label{eqn:lplimiter}
\lambda(R)={{2+R}\over{6+3R+R^2}}.
\end{equation}
Physically, the derivation of this result follows an analysis similar
to the Chapman-Enskog approach for developing solutions to the
Boltzmann equation in the kinetic theory of gases.  In a second
example of a choice of constraint on the anisotropy of the radiation
field, Minerbo (1978) assumed a piecewise linear variation of the
specific intensity with angle, and found the flux limiter
\begin{equation}\label{eqn:minerbolimiter}
\lambda(R)=
  \left\{
    \begin{array}{ll}
      2/(3+\sqrt{9+12R^2})   & \mbox{if $0\leq R\leq 3/2$} \\
      (1+R+\sqrt{1+2R})^{-1} & \mbox{if $3/2<R<\infty$}
    \end{array}
  \right.
\end{equation}
In the optically thin limit $R\rightarrow\infty$, both the LP and
Minerbo flux limiters give to first order in $R^{-1}$
\begin{equation}
\lim_{R\rightarrow\infty}\lambda(R)={1\over R},
\end{equation}
so the magnitude of the flux approaches $|{\bf F}| = c|\nabla E|/(\chi
R) = c E$, which obeys the causality constraint.  In the optically
thick or diffusion limit $R\rightarrow 0$, both examples give to first
order in $R$
\begin{equation}
\lim_{R\rightarrow 0}\lambda(R)={1\over 3},
\end{equation}
so the flux takes the value given by equation~(\ref{eqn:eddflux}).
For intermediate values of $R$, these two forms for the flux limiter
differ substantially, as shown in figure~\ref{fig:limiters}.  Thus a
major uncertainty in FLD calculations is the appropriate choice for
the form of the flux limiter.  Some alternatives are described by
Levermore~(1984).  The LP form has been used in studies of accretion
onto black holes (Eggum, Coroniti, \& Katz 1988) and protostellar
collapse (Bodenheimer et al. 1990).  Dynamics of radiation-dominated
accretion disks computed using the LP and Minerbo forms were compared
by Turner \& Stone~(2001).  Kley~(1989) derived a flux limiter
appropriate for accretion disk boundary layers, using exact solutions
of the transfer equation for spherically-symmetric stellar
atmospheres.  This limiter is quite similar to the Minerbo form over
the whole range of $R$.  A flux limiter appropriate for a scattering
medium was derived and used in accretion disk corona models by Melia
\& Zylstra (1991).  The LP and Minerbo flux limiters have been widely
used, and bracket the range of values of most other flux limiters.
Both are available in the ZEUS-2D implementation of FLD described
here.  Since the appropriate form of the flux limiter is not always
clear beforehand, in some applications new limiters may have to be
derived in order to achieve accurate results.  Comparisons between FLD
and full transport results remain important.

In summary, applying the FLD approximation involves computing the flux
limiter from the value of $R$ in each zone, and then using
equations~(\ref{eqn:fick}) to~(\ref{eqn:flambda}) to obtain the
radiation flux and radiation pressure from the radiation energy
density.

\section{NUMERICAL METHOD\label{sec:method}}

As with the MHD equations in ZEUS-2D, the dynamical equations for the
radiation are operator-split into source and transport terms.  In the
source step, the radiation and material energy densities are updated
using finite difference approximations to
\begin{equation}\label{eqn:erdivflux}
{\partial E\over\partial t} = -\nabla\cdot{\bf F}
\end{equation}
and separately,
\begin{equation}\label{eqn:ersource}
{\partial E\over\partial t} = 
  - \nabla{\bf v}:{\mathsf P} + 4\pi\kappa B - c\kappa E
\end{equation}
and
\begin{equation}\label{eqn:esource}
{\partial e\over\partial t} =
  - p\nabla\cdot{\bf v} - 4\pi\kappa B + c\kappa E.
\end{equation}
In the transport step, an integral formulation is used to generate a
conservative differencing scheme for the advection terms,
\begin{equation}\label{eqn:ertransport}
{d\over dt}\int_V E dV = -\oint_{dV}E({\bf v}-{\bf v_g})\cdot{\bf dS},
\end{equation}
where ${\bf v_g}$ is an arbitrary coordinate velocity which allows
grid motion.  Equations~(\ref{eqn:erdivflux})
to~(\ref{eqn:ertransport}) are evolved in the radiation module, as is
the source term due to radiative acceleration in the gas momentum
equation~(\ref{eqn:gasmomentum}).  The rest of the set of coupled
equations is solved using the existing hydrodynamic algorithms.
Centering of the radiation variables on the hydrodynamic grid is
described in \S\ref{sec:centering}, update of the radiation flux term
in \S\ref{sec:delflux} and the remaining source terms in
\S\ref{sec:eerint}, treatment of the compression terms in the case of
strictly scattering opacity in \S\ref{sec:eerdv}, and the transport
step in \S\ref{sec:transport}.  Stability of the method and the choice
of timestep are discussed in \S\ref{sec:dt}.

\subsection{Centering of Variables\label{sec:centering}}

\begin{figure}
\plotone{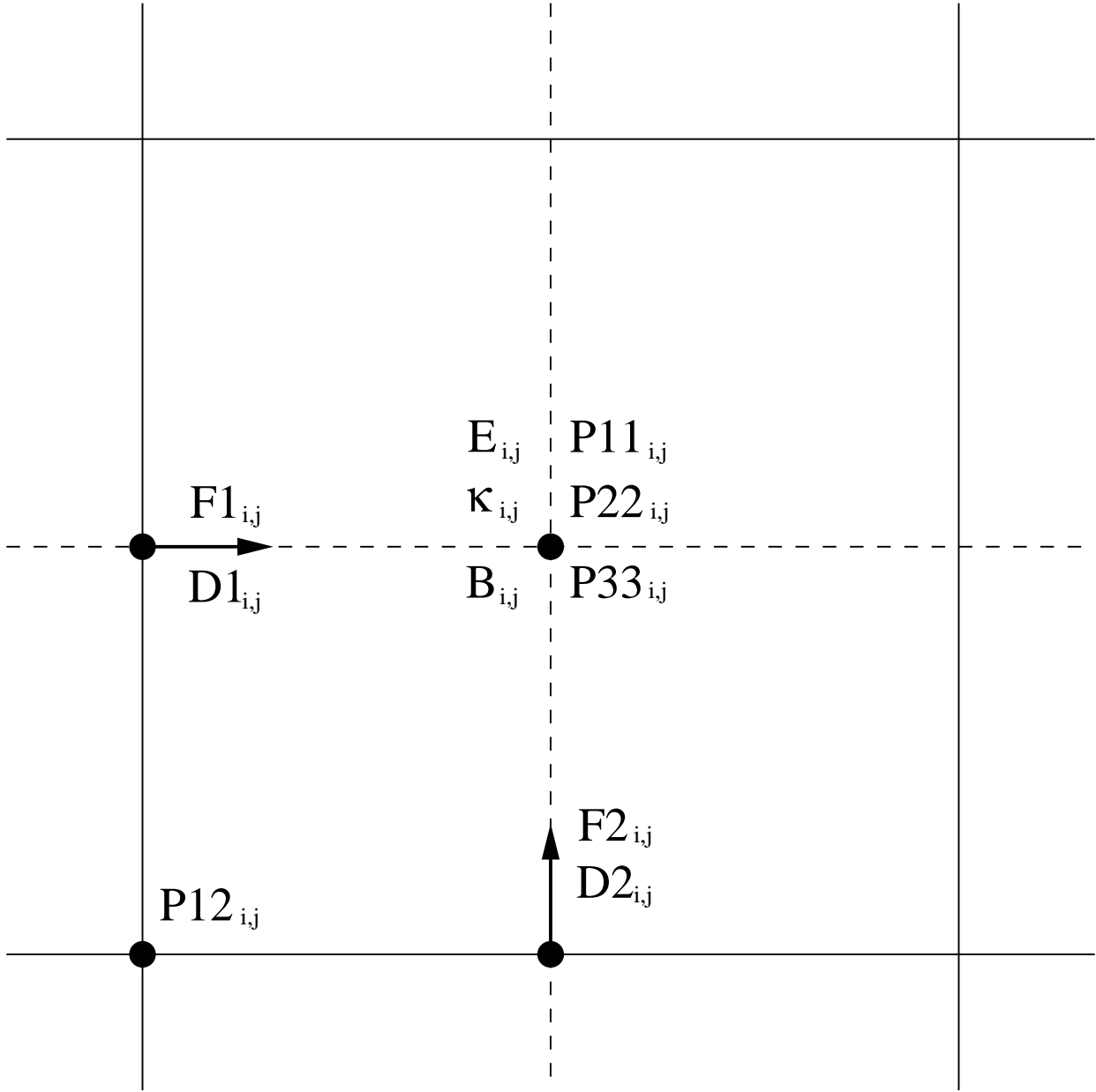}
\caption{\label{fig:mesh}Centering of the radiation hydrodynamic
variables in the ZEUS-2D code.  Symbols are explained in
\S\ref{sec:centering}.  }
\end{figure}

The dependent variables for the radiation are discretized onto the
computational mesh as in ZEUS-2D (Stone \& Norman 1992a, hereafter
SN).  Scalars, and tensors of even rank, are zone-centered, while
tensors of odd rank are face-centered.  The mesh is labeled by
coordinate vectors $x_1$ and $x_2$, with the 3-direction taken to be
orthogonal to the computational plane.  The mesh whose grid lines mark
zone edges is labeled by coordinates $x1a_i$ and $x2a_j$, while the
mesh whose lines intersect at zone centers is labeled by coordinates
$x1b_i$ and $x2b_j$, as shown in figure~\ref{fig:mesh}.  On this mesh,
the radiation variables are discretized according to
\begin{eqnarray*}
     E(x_1,x_2) & \longrightarrow &   E(x1b_i,x2b_j)=  E_{i,j}\\
   F_1(x_1,x_2) & \longrightarrow &  F1(x1a_i,x2b_j)= F1_{i,j}\\
   F_2(x_1,x_2) & \longrightarrow &  F2(x1b_i,x2a_j)= F2_{i,j}\\
P_{11}(x_1,x_2) & \longrightarrow & P11(x1b_i,x2b_j)=P11_{i,j}\\
P_{22}(x_1,x_2) & \longrightarrow & P22(x1b_i,x2b_j)=P22_{i,j}\\
P_{33}(x_1,x_2) & \longrightarrow & P33(x1b_i,x2b_j)=P33_{i,j}\\
P_{12}(x_1,x_2) & \longrightarrow & P12(x1b_i,x2b_j)=P12_{i,j}
\end{eqnarray*}
The components of the Eddington tensor ${\mathsf f}$ are centered
identically to the components of ${\mathsf P}$.  The opacity,
absorption and scattering coefficients, and Planck function are all
zone-centered.  The radiation diffusion coefficient $D$ is computed on
zone 1-faces ($D1_{i,j}$) and 2-faces ($D2_{i,j}$) along with the
components of the flux.

\subsection{Radiation Flux Divergence Term\label{sec:delflux}}

Under the FLD approximation, the flux divergence term in the radiation
energy equation~(\ref{eqn:erdivflux}) is a function of the gradient in
radiation energy density.  When implicitly differenced, it therefore
depends on the time-advanced values of $E$ in adjacent grid zones, and
may be updated by a matrix inversion.  To simplify the form of the
matrix, we operator-split this from the remaining source terms, and
integrate
\begin{equation}\label{eqn:raddiffusion}
{\partial E\over\partial t} = \nabla\cdot(D\nabla E)
\end{equation}
with $D$ given by equation~(\ref{eqn:dc}).  The solution method is
described here for a uniformly-spaced square Cartesian grid.
Extension to the curvilinear coordinates used in ZEUS-2D is
straightforward.  The implicit difference equations to be solved are
of the form
\begin{eqnarray}\label{eqn:delfluxdifferencing}
{(\Delta x)^2\over\Delta t}(E^{n+1}_{i,j}-E^n_{i,j}) &=&
  D1^n_{i+1,j} (E^{n+1}_{i+1,j}-E^{n+1}_{i,j}) -
  D1^n_{i,j}   (E^{n+1}_{i,j}-E^{n+1}_{i-1,j}) +\nonumber \\
&&D2^n_{i,j+1} (E^{n+1}_{i,j+1}-E^{n+1}_{i,j}) -
  D2^n_{i,j}   (E^{n+1}_{i,j}-E^{n+1}_{i,j-1}),
\end{eqnarray}
where $\Delta x$ is the grid spacing, and the spatial indices span
ranges $i=1\ldots N$ and $j=1\ldots M$.  Radiation energy densities
are to be determined at timestep $n+1$.  The difference equations are
kept linear by using diffusion coefficients from step $n$.  This is a
good approximation when $|\nabla E|/(\chi E)$ changes little each
timestep, but may be less appropriate when the radiation field varies
rapidly with time in regions of intermediate or low optical depth.  In
section~\ref{sec:front} it is shown that a propagating radiation front
can be more accurately evolved in this situation when the diffusion
coefficients are recalculated every diffusion substep.

Considering $E^{n+1}_{i,j}$ as a single vector of length $N\times M$,
the updated values may be determined by solving an equation with a
matrix of $(N\times M)^2$ elements, using a sparse-matrix technique
such as incomplete Cholesky -- conjugate gradient (Press et al. 1992).
These methods have proven useful for particular problems with fixed
boundary conditions, but the form of the matrix and the sparse-matrix
method required vary with the boundary conditions.  Solutions of
equation~(\ref{eqn:delfluxdifferencing}) may also be obtained by
combining simple solvers with multi-grid acceleration techniques
(Hackbusch 1985).  However, here we choose an
alternating-direction-implicit (ADI) method on a single grid.  This
may be less efficient than the best sparse-matrix and multi-grid
algorithms, though the number of operations required is proportional
to $N\times M$.  In ADI, successive approximations to $E^{n+1}$ are
computed by advancing towards a $w$-stationary state the equation
\begin{equation}
{\partial E^{n+1}_{i,j}\over\partial w} = - {E^{n+1}_{i,j}-E^n_{i,j}
\over \Delta t} + \left[\nabla\cdot(D\nabla E)\right]^{n+1}_{i,j}.
\end{equation}
The new variable $w$ may be thought of as the pseudo-time.  Each
$w$-step is split into two parts.  In the first, the update is
$w$-implicit along the 1-direction, and explicit along the
2-direction.  In the second, the differencing schemes for the two axes
are exchanged.  Labeling the approximate value of $E^{n+1}_{i,j}$ at
pseudo-timestep $m$ by $E^m_{i,j}$, the difference equations solved
are
\begin{eqnarray}
E^{m+\half}_{i,j}-E^m_{i,j} &=& {\Delta w\over 2\Delta t}
(E^n_{i,j}-E^{m+\half}_{i,j}) + {\Delta w\over 2(\Delta x)^2}\times\mbox{}
\nonumber \\
&&\left\{
  D1^n_{i+1,j} (E^{m+\half}_{i+1,j}-E^{m+\half}_{i,j}) -
  D1^n_{i,j}   (E^{m+\half}_{i,j}-E^{m+\half}_{i-1,j}) \right.
\nonumber\\
&& \left.\mbox{} +
  D2^n_{i,j+1} (E^m_{i,j+1}-E^m_{i,j}) -
  D2^n_{i,j}   (E^m_{i,j}-E^m_{i,j-1})
\right\}
\end{eqnarray}
on sweeps which are implicit along the 1-direction, and
\begin{eqnarray}
E^{m+1}_{i,j}-E^{m+\half}_{i,j} &=& {\Delta w\over 2\Delta t}
(E^n_{i,j}-E^{m+1}_{i,j}) + {\Delta w\over 2(\Delta x)^2}\times\mbox{}
\nonumber \\
&&\left\{
  D1^n_{i+1,j} (E^{m+\half}_{i+1,j}-E^{m+\half}_{i,j}) -
  D1^n_{i,j}   (E^{m+\half}_{i,j}-E^{m+\half}_{i-1,j}) \right.
\nonumber\\
&&\left.\mbox{} +
  D2^n_{i,j+1} (E^{m+1}_{i,j+1}-E^{m+1}_{i,j}) -
  D2^n_{i,j}   (E^{m+1}_{i,j}-E^{m+1}_{i,j-1})
\right\}
\end{eqnarray}
on sweeps which are implicit along the 2-direction.  Each sweep
involves solving a tridiagonal matrix equation.  For example, for
boundary condition $\nabla E=0$, the matrix equation solved on the
$j$th sweep implicit along the 1-direction is
\begin{eqnarray}\label{eqn:tridiagonal}
\left[
\begin{array}{ccccc}
\beta-h D1_{1,j} & -h D1_{2,j} &	       &	     &			  \\
-h D1_{2,j}	 & \beta       & -h D1_{3,j}   &	     &			  \\
		 & -h D1_{3,j} & \beta         & -h D1_{4,j} &                    \\
		 & \ddots      & \ddots        & \ddots      &                    \\
		 &	       & -h D1_{N-1,j} & \beta       & -h D1_{N,j}        \\
		 &	       &	       & -h D1_{N,j} &\beta-h D1_{N+1,j}  \\
\end{array}
\right]
\left[
\begin{array}{c}
E_{1,j} \\
\vdots \\
E_{N,j}
\end{array}
\right]^{m+\half} = {\bf b},
\end{eqnarray}
where $h={\Delta w\over 2(\Delta x)^2}$, $\beta=1+{\Delta w\over
2\Delta t}+h(D1_{i+1,j}+D1_{i,j})$, and the elements of vector ${\bf
b}$ are
\begin{equation}
b_i = [1-h(D2_{i,j+1}+D2_{i,j})] E^m_{i,j} + h D2_{i,j+1} E^m_{i,j+1}
+ h D2_{i,j} E^m_{i,j-1} + {\Delta w\over 2\Delta t} E^n_{i,j}.
\end{equation}
Advancing a pseudo-timestep involves solving $M$ such equations in
$N\times N$ matrices for the 1-implicit sweeps, and $N$ similar
$M\times M$ matrix equations for the 2-implicit sweeps.  A standard
method involving $LU$ decomposition and forward- and back-substitution
is used.  In the case of periodic boundaries, the lower left and upper
right corner elements of the matrices are non-zero, and this condition
is dealt with using the Sherman-Morrison formula (Press et al. 1992)
to compute a correction to the solution of the tridiagonal part.  The
diffusion coefficients are held fixed during the entire real timestep.
Numerical accuracy of the solution is improved by scaling the lengths,
times, and diffusion coefficients in equation~(\ref{eqn:tridiagonal})
by a common factor chosen so the diagonal matrix elements are near
unity.

To bring the solution into pseudo-time steady-state rapidly on
different size scales, the length of the pseudo-timestep $\Delta w$ is
increased exponentially with iteration number $m=1\ldots W$, so that
\begin{equation}
\Delta w = \Delta w_0\left(\Delta w_1\over\Delta w_0\right)^{m-1\over W-1},
\end{equation}
where the initial pseudo-timestep $\Delta w_0$ is one-quarter of the
square of the grid spacing, and the final pseudo-timestep $\Delta w_1$
is one-quarter of the square of the total grid size (Black \&
Bodenheimer 1975).  In cases where the number of zones is large, the
rescaling of lengths and times applied to
equation~(\ref{eqn:tridiagonal}) may not be sufficient to keep the
condition numbers of the matrices near unity for all pseudo-timesteps,
and the reduced accuracy of solutions obtained using $LU$
decomposition may lead to slow convergence.  Under these conditions it
might be helpful to iteratively improve the initial solution, or to
use an alternative decomposition such as $QR$ or Householder (Press et
al. 1992).  For the test problems discussed in
section~\ref{sec:tests}, reducing the condition numbers of the
matrices by adjusting the rescaling factor did not speed convergence.

The error in the time-advanced solution $E^{n+1}$ obtained from the
series of $w$-steps is checked by substituting in the original
difference equation~(\ref{eqn:delfluxdifferencing}).  The residual is
normalized against the ratio of the time-centered $E$ to the
hydrodynamical timestep.  If the biggest error on the grid is too
large, the method is applied again with larger $W$ and more
closely-spaced pseudo-timesteps.  If a sufficiently small error is not
reached after many pseudo-timesteps, then the real timestep for the
flux divergence term is halved, and the term is applied twice per
hydrodynamical timestep.  If a sufficiently small error is not
obtained with the real timestep finely divided, the calculation is
terminated.  On the other hand, if the error obtained is very small,
then fewer pseudo-timesteps are tried initially in the following
diffusion substep.

\subsection{Matter-Radiation Interaction and Compressive Heating Terms
\label{sec:eerint}}

The absorption, emission, and compressive heating source terms are
approximated by implicit difference equations, to obtain numerical
stability for timesteps which may be long compared with the
matter-radiation interaction timescale.  The solution is advanced in
these terms from step $n$ to step $n+1$ by solving simultaneously
\begin{equation}\label{eqn:erdifference}
E^{n+1}_{i,j}-E^n_{i,j} = \Delta t\left\{-(\nabla{\bf v}:{\mathsf
P})^{n+1}_{i,j} + 4\pi\kappa B^{n+1}_{i,j} - c\kappa
E^{n+1}_{i,j}\right\}
\end{equation}
and
\begin{equation}\label{eqn:edifference}
e^{n+1}_{i,j}-e^n_{i,j} = \Delta t \left\{ -p^{n+1}_{i,j}
(\nabla\cdot{\bf v})_{i,j} - 4\pi\kappa B^{n+1}_{i,j} + c\kappa
E^{n+1}_{i,j} \right\}.
\end{equation}
These depend on the values of $e$ and $E$ in zone $i,j$ only, so the
spatial coordinate subscripts are dropped in subsequent equations.
The gas is assumed to be ideal, with ratio of specific heats $\gamma$,
and equation of state $p^{n+1}=(\gamma-1)e^{n+1}$.  The Planck
function $B^{n+1}={1\over\pi}\sigma_B T^4$ is computed from the
temperature $T=(\gamma-1)\mu e^{n+1}/({\cal R}\rho^n)$ using the
density $\rho^n$ at the preceding time.  Here $\mu$ is the
dimensionless mean particle mass, ${\cal R}$ the gas constant, and
$\sigma_B$ the Boltzmann constant.  The rate $(\nabla{\bf v}:{\mathsf
P})^{n+1}$ at which work is done on the radiation by the flow is
\begin{equation}\label{eqn:gradvprad}
\left\{\nabla{\bf v}_{11}f11 + \nabla{\bf v}_{22} f22 + \nabla{\bf
v}_{33}(1-f11-f22) + (\left\langle\nabla{\bf v}_{12}\right\rangle +
\left\langle\nabla{\bf v}_{21}\right\rangle) f12 \right\} E^{n+1}.
\end{equation}
The Eddington tensor components $f11$, $f22$, and $f12$, the fluid
velocity ${\bf v}$, and the absorption opacity $\kappa$ are evaluated
at timestep $n$, and the components of the velocity gradient tensor
are as in Stone, Mihalas, \& Norman (1992) Appendix~A.  With these
choices, the only unknowns in equations~(\ref{eqn:erdifference})
and~(\ref{eqn:edifference}) are $e^{n+1}$ and $E^{n+1}$.  Eliminating
$E^{n+1}$ between the two equations yields a quartic polynomial with
$e^{n+1}$ as one root.  With $x$ in place of $e^{n+1}$, the quartic is
\begin{equation}\label{eqn:interactionquartic}
x^4 + {(1+a_4)(1+a_2+a_3)\over a_1(1+a_3)} x - {(1+a_2+a_3)e^n + a_2
E^n \over a_1(1+a_3)} = 0,
\end{equation}
where
\begin{equation}
a_1=4\kappa\sigma_B\left\{{\mu(\gamma-1)\over{\cal R}\rho^n}\right\}^4
\Delta t,
\end{equation}
\begin{equation}
a_2=c\kappa\Delta t,
\end{equation}
\begin{equation}
a_3={(\nabla{\bf v}:{\mathsf P})^{n+1} \over E^{n+1}}\Delta t,
\end{equation}
and
\begin{equation}
a_4=(\gamma-1)(\nabla\cdot{\bf v})\Delta t.
\end{equation}
The quartic typically has a single positive real root when the
timestep satisfies the Courant condition of \S\ref{sec:dt}.  Writing
the coefficient of $x$ in equation~(\ref{eqn:interactionquartic}) as
$c_1$ and the constant coefficient as $c_0$, this single root lies
between zero, and the smaller of $|c_0/c_1|$ and $|c_0|^{1/4}$.  The
updated gas energy density is obtained by finding the root using
Newton-Raphson iteration on this interval, with bisection when the
Newton-Raphson method fails.  The updated radiation energy density is
determined by substituting the updated gas energy density in
equation~(\ref{eqn:erdifference}).

\subsection{The Case of Zero Absorption Opacity\label{sec:eerdv}}

When the absorption opacity $\kappa$ is zero and the
material-radiation interaction terms vanish,
equations~(\ref{eqn:erdifference}) and~(\ref{eqn:edifference}) are no
longer coupled.  In this instance, numerical energy conservation may
be improved by using time-centered values for the gas and radiation
pressures.  In this differencing scheme, the time-advanced gas energy
density $e$ is
\begin{equation}
e^{n+1} = e^n \left[{1-\frac{1}{2}q\Delta t \over 1+\frac{1}{2}q\Delta
t} \right],
\end{equation}
with $q=(\gamma-1)(\nabla\cdot{\bf v})^n$.  For the radiation energy
density update, $q$ is replaced by the quantity multiplying $E^{n+1}$
in equation~(\ref{eqn:gradvprad}).

\subsection{Transport Step\label{sec:transport}}

Advection of the radiation energy proceeds exactly as described by SN
for the hydrodynamic variables.  Having expressed the advection terms
in integral form in equation~(\ref{eqn:ertransport}), we apply the
conservative differencing scheme used for the hydrodynamic variables.
In this scheme, the flux of advected radiation energy is computed from
the mass flux in order to reduce the relative numerical diffusion of
radiation with respect to the gas.  The flux across every zone
interface is computed using an interpolation method which may be
selected as either donor cell, van Leer, or piecewise parabolic.
These fluxes are then used to update the radiation energy density in a
directionally split fashion.

Thus, the flux of radiation energy density along the 1-direction is
constructed using
\begin{equation}
{\cal F}^1_{i,j} = (E/d)^*_{1,i,j} {\dot M}^1_{i,j} g2a_i^{n+\half}
g31a_i^{n+\half} dvl2a_j^n,
\end{equation}
where the mass fluxes in the 1-direction, ${\dot M}^1_{i,j}$, are
defined by equation [55] in SN.  These fluxes are used to update the
radiation energy under 1-transport via
\begin{equation}
E_{i,j}^{n+1} dvl1a_i^{n+1} dvl2a_j^n - E_{i,j}^n dvl1a_i^n dvl2a_j^n
= -\Delta t({\cal F}^1_{i,j+1} - {\cal F}^1_{i,j}).
\end{equation}
Advection fluxes of radiation energy density in the 2-direction are
computed by
\begin{equation}
{\cal F}^2_{i,j} = (E/d)^*_{2,i,j} {\dot M}^2_{i,j} g31b_i^n dx1b_i^n
g32a_j^{n+\half},
\end{equation}
where the mass fluxes in the 2-direction, ${\dot M}^2_{i,j}$, are
defined by equation [56] in SN.  These fluxes are then used to update
the radiation energy due to advection in the 2-direction via
\begin{equation}
E_{i,j}^{n+1} dvl1a_i^n dvl2a_j^{n+1}-E_{i,j}^n dvl1a_i^n dvl2a_j^n =
-\Delta t({\cal F}^2_{i,j+1}-{\cal F}^2_{i,j}).
\end{equation}
To reduce the systematic effects of the directional splitting
discussed in section~\ref{sec:testtransport}, the order in which the
two advection substeps are applied is reversed each timestep.

\subsection{Stability and the Timestep Criterion\label{sec:dt}}

Since the advection terms in the RHD equations are treated
time-explicitly, the calculations may be numerically unstable unless
the timestep is less than the time for waves or advection to carry
energy across a grid zone (Richtmyer \& Morton 1957).  The wave family
with the largest group speed may be either adiabatic sound waves or
radiation acoustic waves, depending on the ratio of gas and radiation
energy densities.  The sound speed for purposes of computing the
timestep is therefore chosen to be
\begin{equation}
c_s = \left[\max(\gamma, \frac{4}{3}) {P_{tot}\over\rho} \right]^{1/2},
\end{equation}
where $P_{tot}$ is the sum of the gas pressure and the largest
component of the radiation pressure tensor.  The remainder of the
calculation of the timestep proceeds as in the hydrodynamic portion of
ZEUS-2D (SN).  Though resulting steps may be longer than the time for
radiation to diffuse across a zone $(\Delta x)^2\over D$, the implicit
differencing of the radiation diffusion term (\S\ref{sec:delflux})
ensures stability of the method.

\section{TEST CALCULATIONS\label{sec:tests}}

The problems used to test the radiation module fall into three
categories.  Calculations outlined in the first three sections below
test in isolation the heating and cooling terms
(\S\ref{sec:testheatcool}), the radiation flux divergence term
(\S\ref{sec:testdelflux}), and the transport terms
(\S\ref{sec:testtransport}).  Linear RHD wave tests involving all the
radiation terms are described in \S\ref{sec:linear}.  Calculations
which exercise radiation terms in the non-linear regime are
propagation of optically-thin radiation fronts (\S\ref{sec:front}),
structure of radiating gas pressure dominated shocks of moderate
optical depth (\S\ref{sec:critical}), and formation of a steady
optically-thick shock in which radiation pressure dominates downstream
(\S\ref{sec:raddomshock}).

\subsection{Heating and Cooling Terms \label{sec:testheatcool}}

\begin{figure}
\plotone{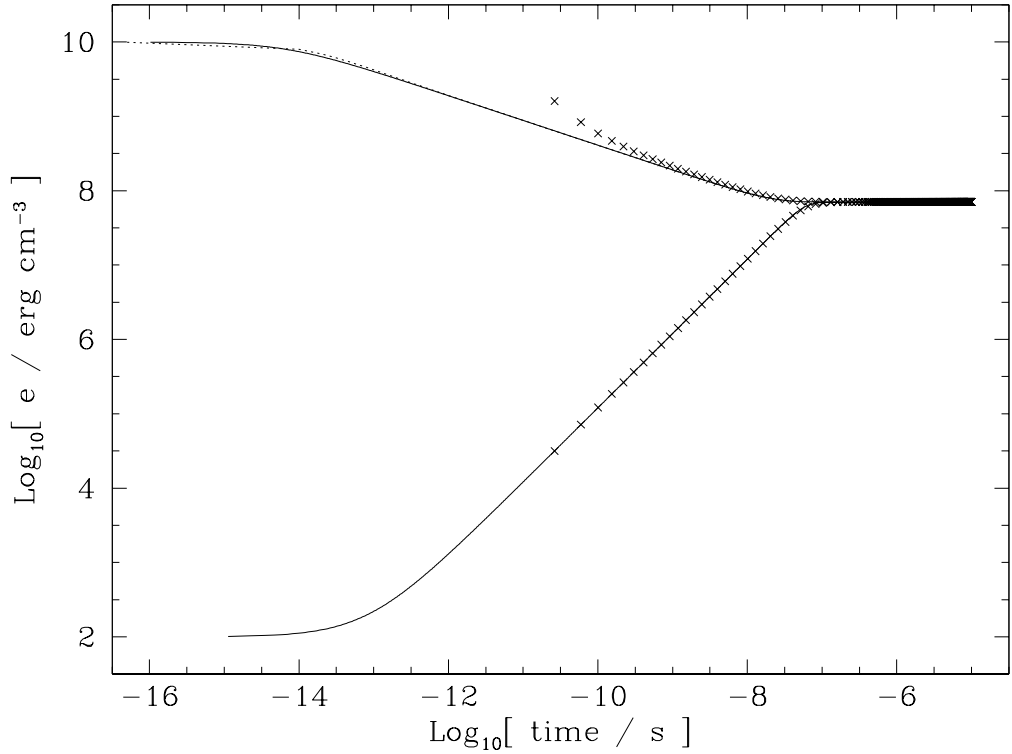}
\caption{\label{fig:testheatcool}Results from calculations of approach
to thermal equilibrium are shown by crosses every timestep.  Starting
value of the thermal energy density is $10^{10}$~erg~cm$^{-3}$ for the
upper set of crosses, and $10^2$~erg~cm$^{-3}$ for the lower set.  The
initial radiation energy density $E$ is $10^{12}$~erg~cm$^{-3}$ in
both cases.  Corresponding analytic solutions assuming constant $E$
are indicated by solid curves.  Results from a third calculation, with
initial energy densities matching the upper case, and timestep fixed
at $10^{-14}$~seconds, are shown by a dotted line almost coincident
with the upper solid curve.  }
\end{figure}

In a static uniform absorbing fluid initially out of thermal balance,
an analytic solution for the time evolution of the gas energy may be
obtained for the case where radiation energy dominates the total.  The
radiation energy emitted or absorbed in reaching equilibrium is a
small fraction of the initial value, so radiation energy density $E$
is assumed constant.  The time evolution of the gas energy density $e$
is obtained by solving the ordinary differential equation
\begin{equation}
{de\over dt} = c \kappa E - 4 \pi \kappa B(e).
\end{equation}
Solutions are plotted together with the numerical results in
figure~\ref{fig:testheatcool}, for a density of $10^{-7}$~g~cm$^{-3}$,
opacity $4\times 10^{-8}$~cm$^{-1}$, mean dimensionless mass per
particle 0.6, and index in the equation of state $\gamma=5/3$.  Two
cases are shown.  In the first, the initial thermal energy density is
less than the equilibrium value.  The agreement of the computed result
with the analytic solution in this case is excellent.  In the second
case, $e$ is initially larger than the equilibrium value, lags the
expected decline for the first few steps, and reaches equilibrium in
the correct time.  These results are consistent with the use of a
fully-implicit method and a timestep of a few times
$10^{-11}$~seconds, much longer than the cooling time.  When the
timestep is held to less than the cooling time, the agreement between
the calculation and the analytic solution is improved, as shown by the
dotted line in figure~\ref{fig:testheatcool}.  The sum of thermal and
radiation energies is conserved in these three cases to about the
floating-point precision.

\subsection{Flux Divergence Term \label{sec:testdelflux}}

An analytic solution of the diffusion
equation~(\ref{eqn:raddiffusion}) on the unit square, with periodic
boundaries and unit diffusion coefficient, is (Lamb~1995)
\begin{equation}
E(x,y,t) = 2 + e^{-8\pi^2 t} \sin{2\pi x} \sin{2\pi y}.
\end{equation}
This function at $t=0$ is discretized onto a grid with 100 zones along
each side, and the evolution followed using the radiation diffusion
term alone.  The timestep chosen is a factor of a hundred longer than
the time $(\Delta x)^2\over D$ for radiation to diffuse across a zone.
The largest zone-by-zone difference between the numerical and analytic
solutions during the evolution towards equilibrium is 5.3\%, and the
largest root-mean-square difference is 2.5\%.  When the timestep is
shortened by a factor ten, the corresponding residuals are 0.7\% and
0.3\%.  The implicit numerical solution relaxes towards equilibrium
more slowly than the analytic solution in each case.

\subsection{Transport Terms \label{sec:testtransport}}

When the directionally-split advection substeps described in
section~\ref{sec:transport} are applied in the same order each
timestep, the splitting leads to a systematic delay in transport along
the first axis, and an advance in transport along the second axis.  As
a test of this effect, the source terms were switched off, and a
circular region of increased radiation energy density initially at the
center of the domain was carried twice diagonally across a grid of
$32^2$ zones with periodic boundaries.  About 2000 timesteps were
taken, and the advection substeps were applied always in the order 1,
2.  At the end of this calculation, the centroid of the region lay off
the grid diagonal by 0.02 zones along each axis.  When the calculation
was repeated with the order of the substeps reversed each timestep,
the centroid remained within $3\times 10^{-7}$~zones of the diagonal.
We have chosen to reverse the order of advection each timestep in all
further calculations discussed in this article, in order to minimize
the effects of the directional splitting of the transport terms.

\subsection{Linear RHD Waves \label{sec:linear}}

\begin{figure}
\plottwo{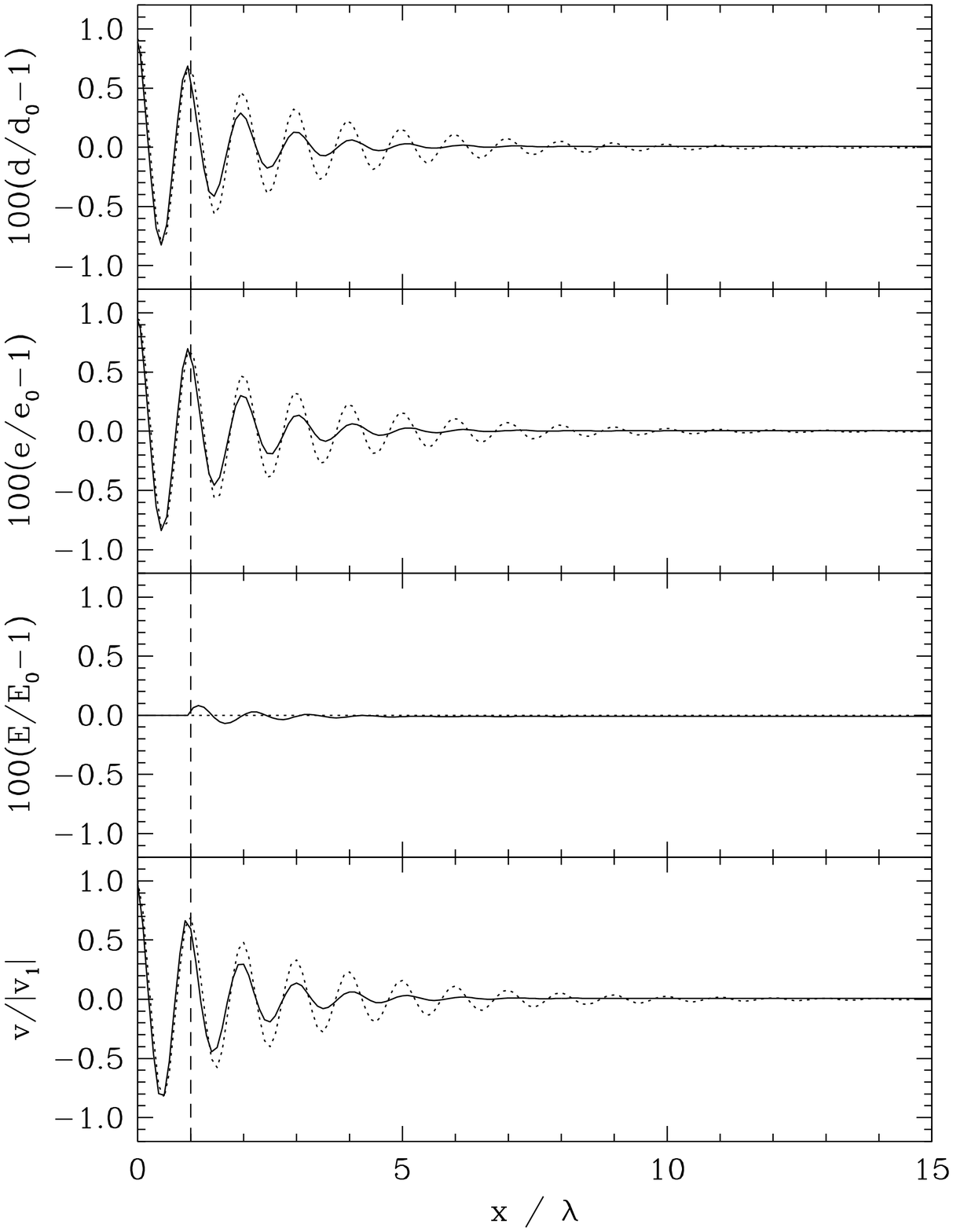}{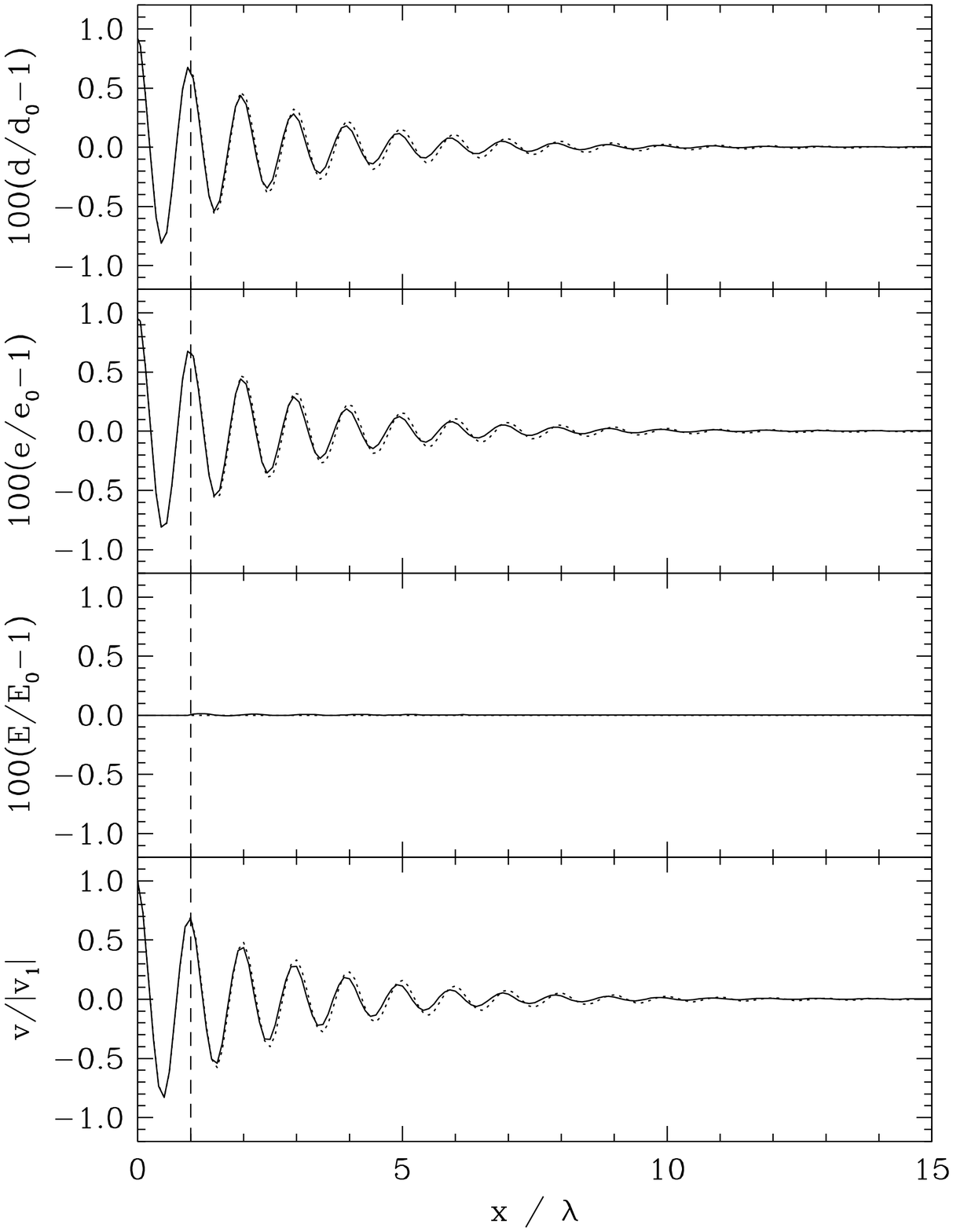}
\caption{\label{fig:taum3}An acoustic oscillation driven in the region
left of the vertical dashed line is damped by radiative cooling as it
travels to the right.  Optical depth per wavelength is $10^{-3}$.
Dotted lines indicate the linear analytic solution, solid lines the
numerical solutions.  The quantities plotted are the relative
perturbations in (top to bottom) density, thermal energy density,
radiation energy density, and velocity.  There are ten grid zones per
wavelength $\lambda$.  For the left panel, the timestep is set by the
Courant condition, and is longer than the time for radiation to
diffuse across a zone by a factor $1.5\times 10^7$.  For the right
panel, the timestep was reduced by a factor ten.  }
\end{figure}

\begin{figure}
\plottwo{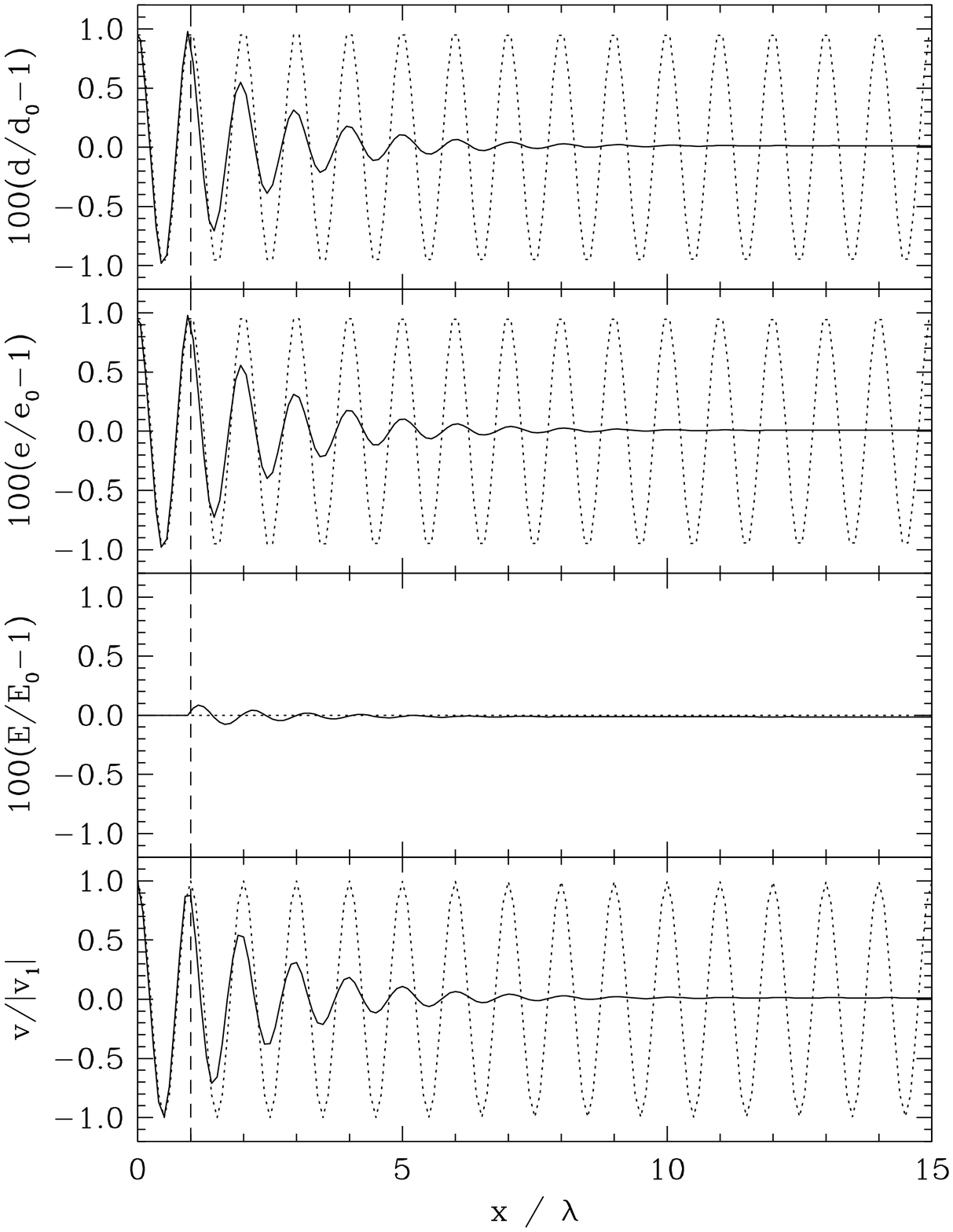}{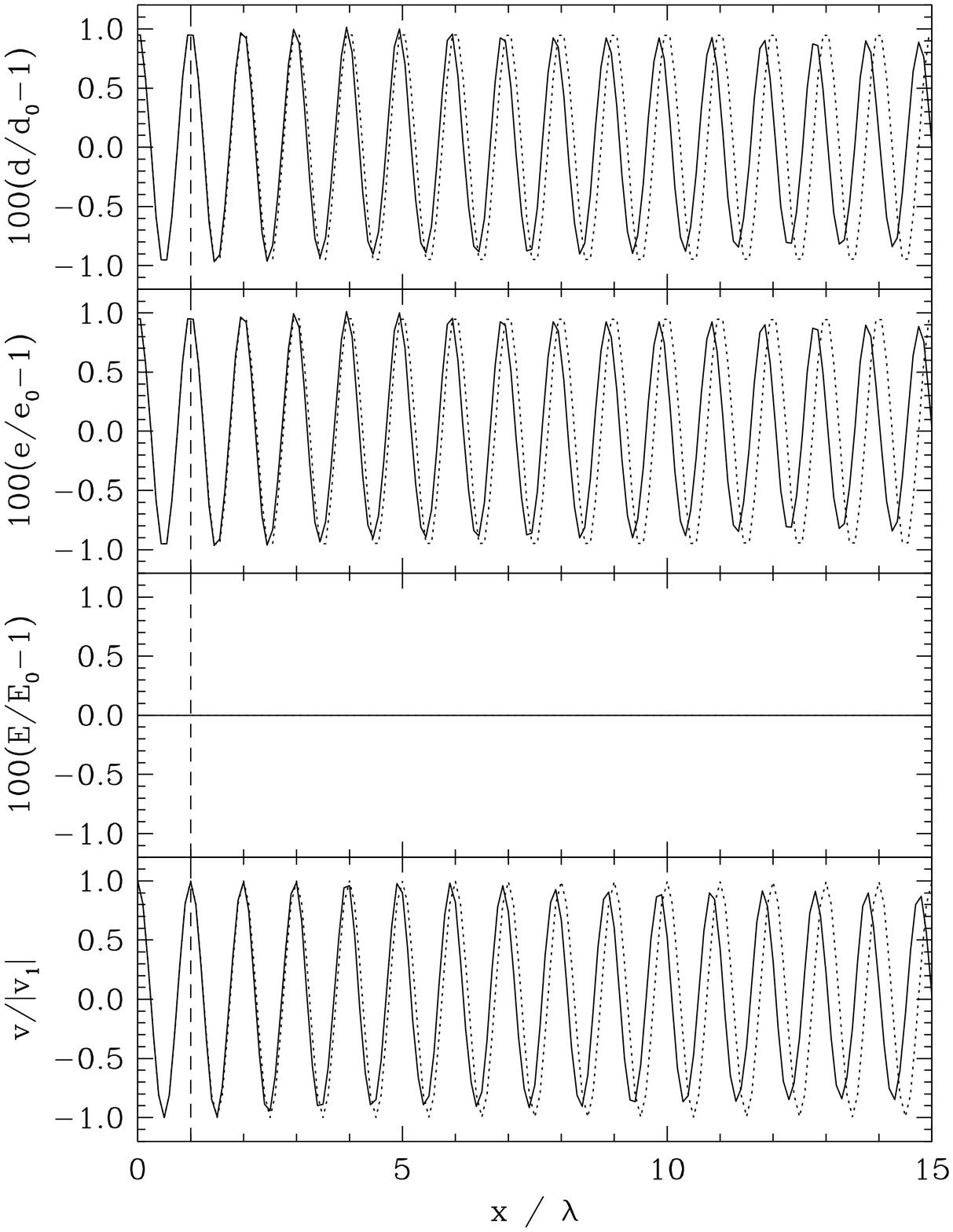}
\caption{\label{fig:tau0}Same as figure~\ref{fig:taum3}, but for a
slower oscillation such that the optical depth per wavelength is
unity, and damping by radiation is slight.  There are ten grid zones
per wavelength $\lambda$.  For the left panel, the timestep is set by
the Courant condition, and is longer than the time for radiation to
diffuse across a zone by a factor $1.5\times 10^4$.  For the right
panel, the timestep was reduced by a factor 1000.  }
\end{figure}

\begin{figure}
\plottwo{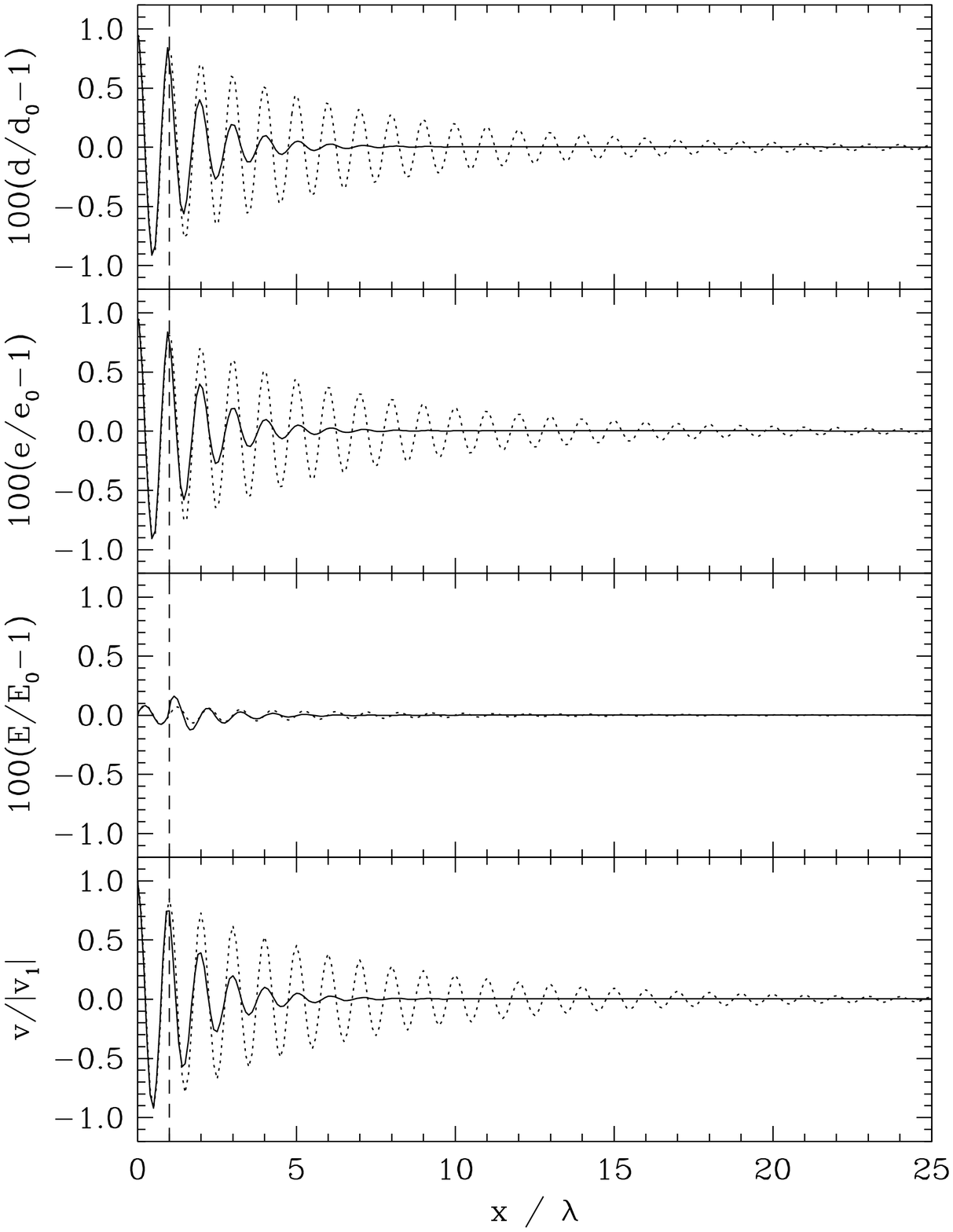}{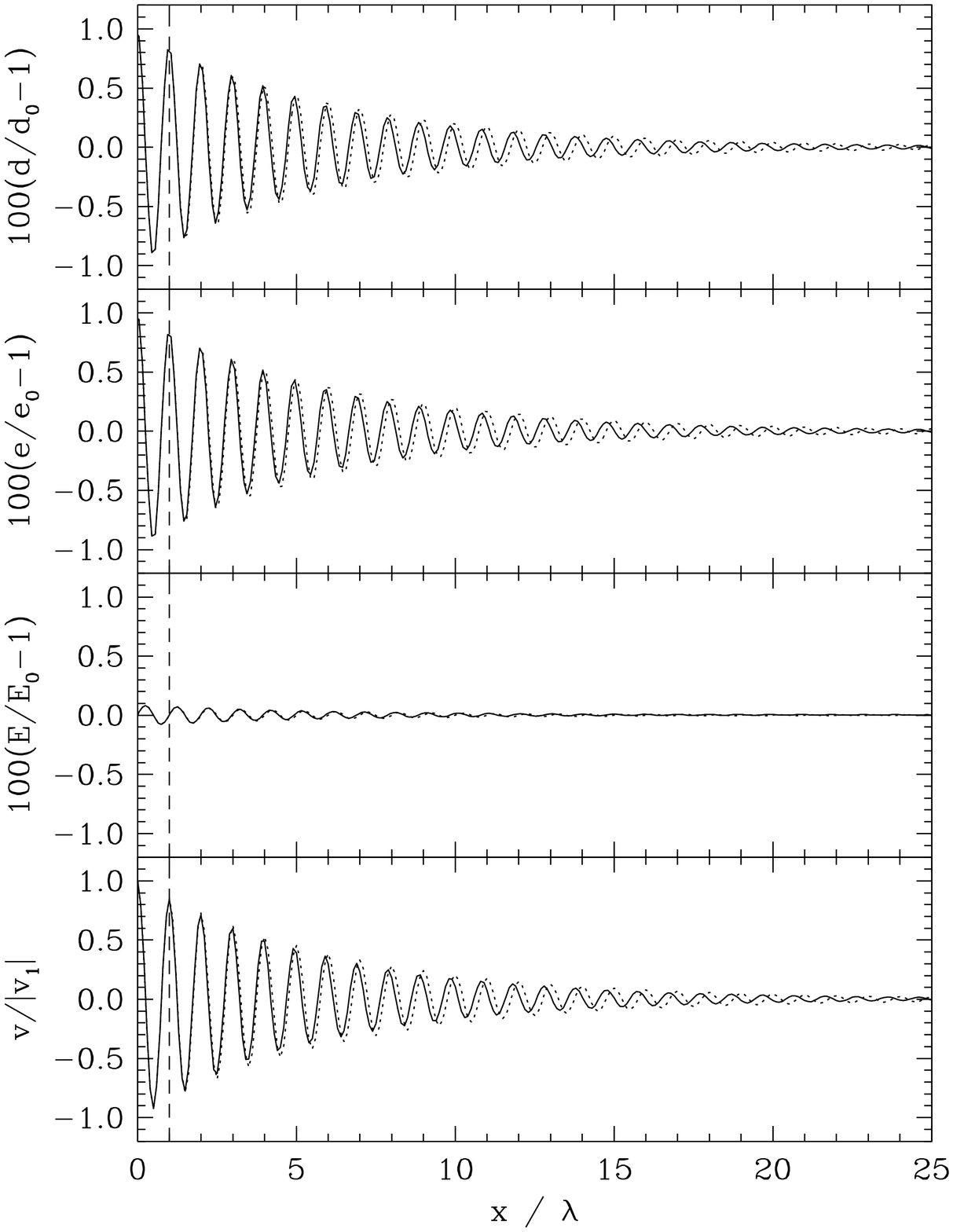}
\caption{\label{fig:taup3}Same as figure~\ref{fig:taum3}, but for a
still slower oscillation such that the optical depth per wavelength is
1000, and damping occurs by radiative diffusion.  There are ten grid
zones per wavelength $\lambda$.  For the left panel, the timestep is
set by the Courant condition, and is longer than the time for
radiation to diffuse across a zone by a factor 15.  For the right
panel, the timestep was reduced by a factor 100.  }
\end{figure}

Waves in a radiating fluid may be damped by escape of photons.
Damping is often rapid for oscillations with frequency near the rate
at which material cools by radiating, and also for oscillations with
wavelength near the distance radiation diffuses in a wave period.
Mihalas \& Mihalas (1983) obtained a linear dispersion relation for
radiation and acoustic plane waves, driven by a boundary into a
uniform medium with opacity due to absorption.  They made the
Eddington approximation, and assumed a uniform background in thermal
equilibrium.  The coefficients of the dispersion relation may be
specified by two dimensionless parameters.  The Boltzmann number
$\mbox{Bo}\equiv{4\gamma c_a e/(c E)}$ is a ratio of typical rates of
energy transport due to material advection and radiation in the
background state, and $r\equiv {E/(4\gamma e)}$ fixes the ratio of the
energy densities.  In these definitions, $\gamma$ is the index in the
equation of state and $c_a$ is the adiabatic sound speed.  For the
tests discussed below, $\mbox{Bo}=10^{-3}$ and $r=0.1$.  Three
frequencies are considered, selected for optical depth per wavelength
$\tau_\lambda=10^{-3}$, unity, and 1000.  The first of these
frequencies is near the radiative cooling rate, and the last mode has
wavelength near the radiative diffusion distance.  In the test
calculations, absorption opacity is set to 0.4~cm$^2$~g$^{-1}$ times
the density.  The domain is initially stationary and uniform, and the
chosen $\mbox{Bo}$ and $r$ correspond to density $3.216\times
10^{-9}$~g~cm$^{-3}$, gas energy density 26020~erg~cm$^{-3}$, and
radiation energy density 17340~erg~cm$^{-3}$.  A region one wavelength
long at the left of the domain is updated each timestep with the time
and space variation found from the linear analysis.  A wave propagates
to the right from this driving region.  Late-time results for waves of
the three optical depths are shown in figures~\ref{fig:taum3}
to~\ref{fig:taup3}.  There are ten zones per wavelength in each case.
Timesteps were chosen according to the Courant condition of
\S\ref{sec:dt} for the calculations in the left-hand panel of each
figure.  The resulting steps, though having Courant numbers of 0.5,
are longer than the time $(\Delta x)^2/D$ for radiation to diffuse
across a zone.  The fully-implicit method used remains stable for
these long timesteps, but loses accuracy, and the waves are damped
more strongly than expected.  Similar damping occurs when the
Eddington approximation is made, indicating the effect is not due to
the explicit differencing of the radiation diffusion coefficient in
equation~(\ref{eqn:delfluxdifferencing}).  In the right-hand panels of
figures~\ref{fig:taum3} to~\ref{fig:taup3} are results from
calculations in which the timesteps were reduced so the method
adequately reproduced the linear analytic results.  The wavelengths
obtained are one to two percent shorter than the predicted
wavelengths.  Similar agreement with the linear predictions was
observed for $\mbox{Bo}=10^{-6}$ and $r=1000$.

\subsection{Propagating Radiation Fronts in Optically Thin Media
\label{sec:front}}

\begin{figure}
\plotone{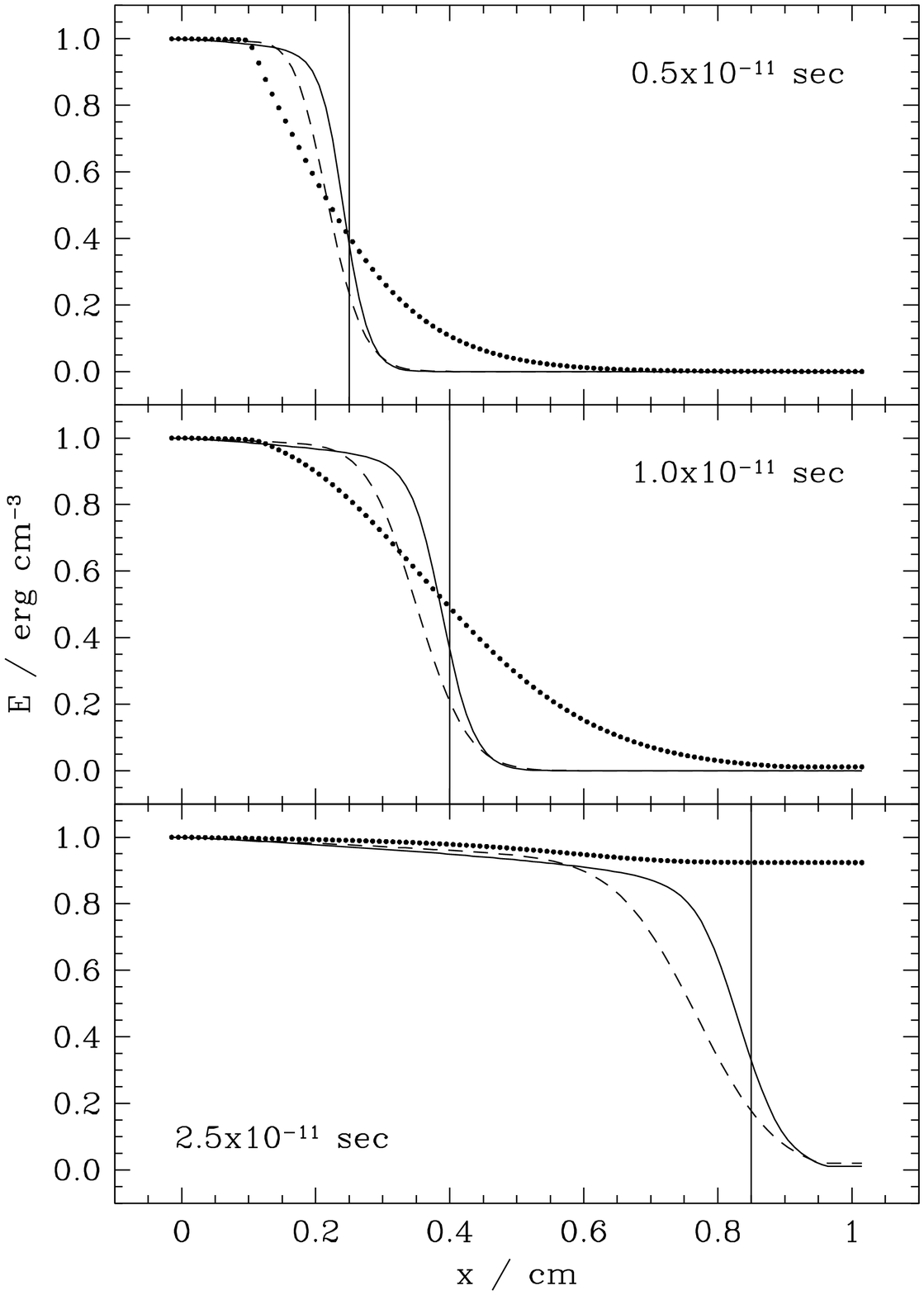}
\caption{\label{fig:front}Radiation front propagating across a box of
optical depth 0.01 is shown at three different times.  The solution
obtained with timestep $\Delta x/(2c)$ is shown by dashed curves.
Results with timestep a factor ten longer are indicated by dots, and
those with timestep a factor ten shorter by solid curves.  Vertical
lines indicate the expected positions of the front.  }
\end{figure}

In optically-thin regions, changes in radiation energy density
propagate at the speed of light.  To examine how well the method
follows such changes, a domain 1~cm long is placed in thermal
equilibrium with a radiation energy density of
$10^{-22}$~erg~cm$^{-3}$.  The mass density is set to
0.025~g~cm$^{-3}$ and the opacity to 0.4~cm$^2$~g$^{-1}$, so optical
depth across the domain is 0.01.  One hundred grid zones are used.  At
time $t=0$, the radiation energy density in zones to the left of
$x=0.1$ is raised to unity, and the increase is allowed to propagate
to the right.  The left-hand boundary condition is inflow, the
right-hand outflow.  Results using the LP flux limiter and scattering
opacity are shown in figure~\ref{fig:front}.  The front moves at
approximately the speed of light.  It is spread over a large number of
zones, indicating the method is quite diffusive in this situation.
For problems in which details of such fronts are important, it may be
better to solve the time-dependent transfer equation using steps set
by the radiation flow time.  With timesteps longer than $\Delta
x/(2c)$, the front is wider still, however the method remains stable.
The width is less when the diffusion coefficients are recalculated
every diffusion substep, rather than once per hydrodynamic timestep.
Where the leading edge of the front reaches the end of the grid, a
region of uniform lower radiation energy density is seen.  The effect
is greatest when the timestep is long and the front is broad, as in
the curve shown by dots on the bottom panel of figure~\ref{fig:front}.
Radiation energy accumulates in this region because the boundary
condition $\nabla E=0$ forces the radiative flux to zero at the
boundary.  Other conditions on radiation energy density which may be
appropriate across outflow boundaries are fixed values of $E$, and
$\nabla\cdot{\bf F}=0$.

\begin{figure}
\plotone{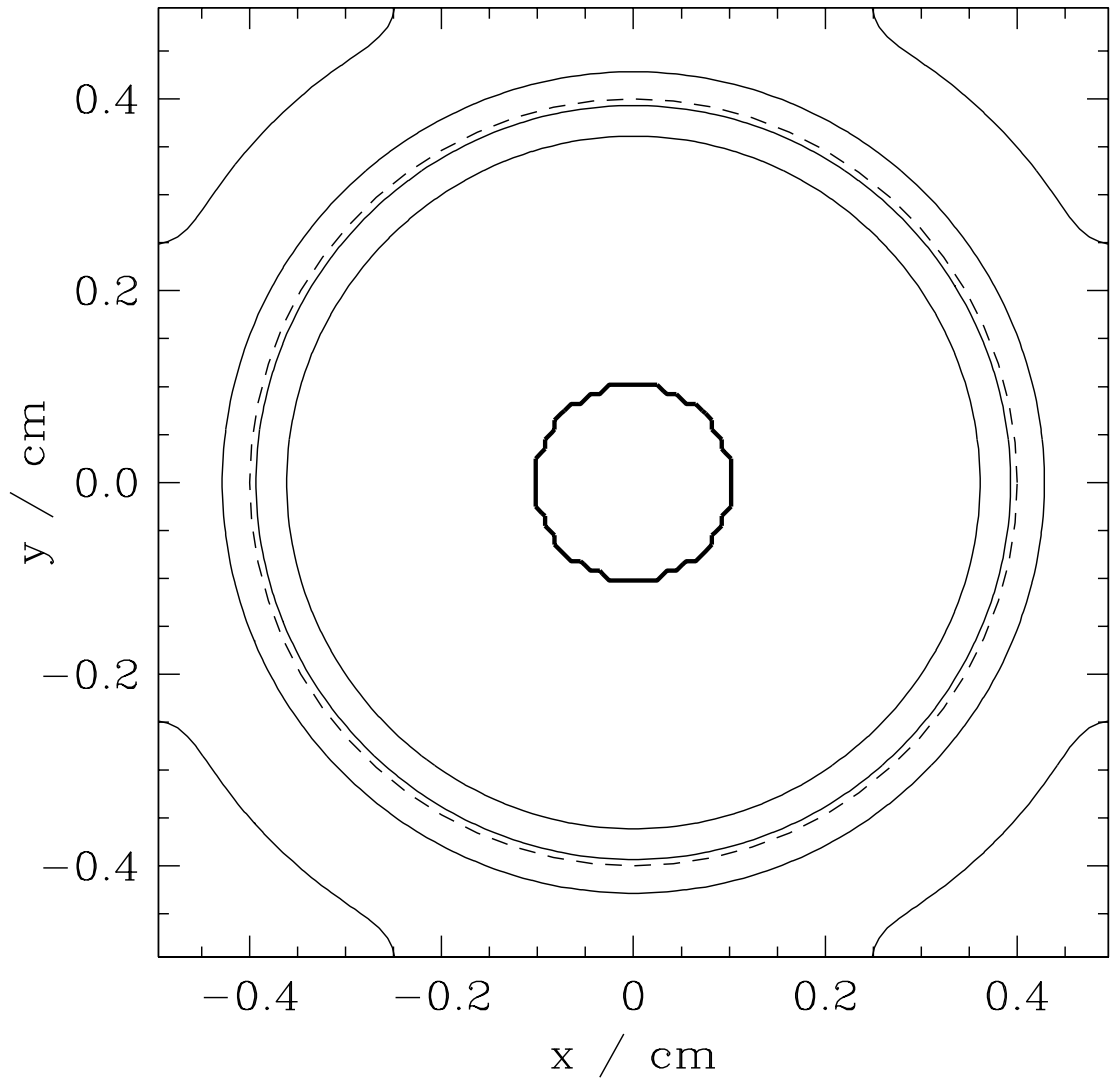}
\caption{\label{fig:front2d}Radiation front expanding from near the
center of a square of optical depth 0.01 per side.  Region where
radiation energy is initially enhanced is marked by a heavy contour.
Expected location of the front after $10^{-11}$~sec is shown by dashed
circle.  Calculated distribution of radiation energy density at this
time is shown by light contours at 1\%, 25\%, 50\%, and 75\% of the
maximum.  }
\end{figure}

The off-diagonal components of the Eddington tensor
(equation~(\ref{eqn:eddfactor})) may be non-zero in radiation fronts
inclined with respect to the coordinate axes.  As a test of the method
in such a situation, the domain used in the one-dimensional radiation
front calculation is expanded to a 1~cm square of $100\times
100$~zones.  The high radiation energy density is placed initially in
zones within 0.1~cm of grid center.  The boundaries are made periodic
in both directions, and the timestep is fixed at $\Delta x/(2c)$.  The
distribution of radiation after $10^{-11}$ seconds is shown in
figure~\ref{fig:front2d}.  The lowest contours on the figure are
non-circular.  After the leading edge of the front reaches the
midpoints of the grid boundaries, the gradient in radiation energy
density there is directed along the boundaries, rather than away from
grid center.  Since in the FLD approximation the flux is assumed
parallel to $-\nabla E$, an incorrect result is obtained in these
regions.  A similar error will result in FLD calculations whenever
inclined radiation fronts meet.  After the front has passed the
corners of the grid, at $5\times 10^{-11}$~sec, total energy differs
from that at the start of the calculation by about one part in
$10^{14}$, indicating that energy is conserved to high precision in
this instance.

\subsection{Subcritical and Supercritical Shocks \label{sec:critical}}

The structure of a shock in radiating fluid is sometimes altered by
photons propagating upstream to preheat approaching material.  If
material reaching the shock front remains cooler than the shocked gas,
the shock is termed subcritical, and the main effect is a slightly
raised post-shock temperature, which declines downstream as radiation
is emitted upstream.  In stronger, supercritical shocks, upstream
material is heated almost to the temperature of the post-shock gas,
and material immediately upstream is in thermal equilibrium with the
radiation emitted through the front.  Gas reaching the front then
overshoots the final temperature significantly, and cools to the
post-shock temperature by radiating from a layer with optical depth
less than unity.  In this section we compare structures calculated for
subcritical and supercritical shocks against analytic estimates by
Zel'dovich \& Raizer (1967), and numerical results obtained by
Sincell, Gehmeyr, \& Mihalas (1999) using a radiation hydrodynamics
code with adaptive refinement of a one-dimensional grid.  The problem
is set up as by Sincell et al. (1999).  The gas is given absorption
opacity 0.4~cm$^2$~g$^{-1}$, mean molecular weight 0.5, and uniform
density $7.78\times 10^{-10}$~g~cm$^{-3}$.  Gas temperature
$10+75x/(7\times 10^{10}$~cm$)$~K increases linearly with distance $x$
from the origin, and the radiation energy density is chosen for
equilibrium.  Grid spacing is set to $1.4\times 10^8$~cm, so initially
the optical depth per zone is 0.044.  A piston moves into the gas from
the origin at uniform speed, and shocked material accumulates in front
of the piston.

\begin{figure}
\plotone{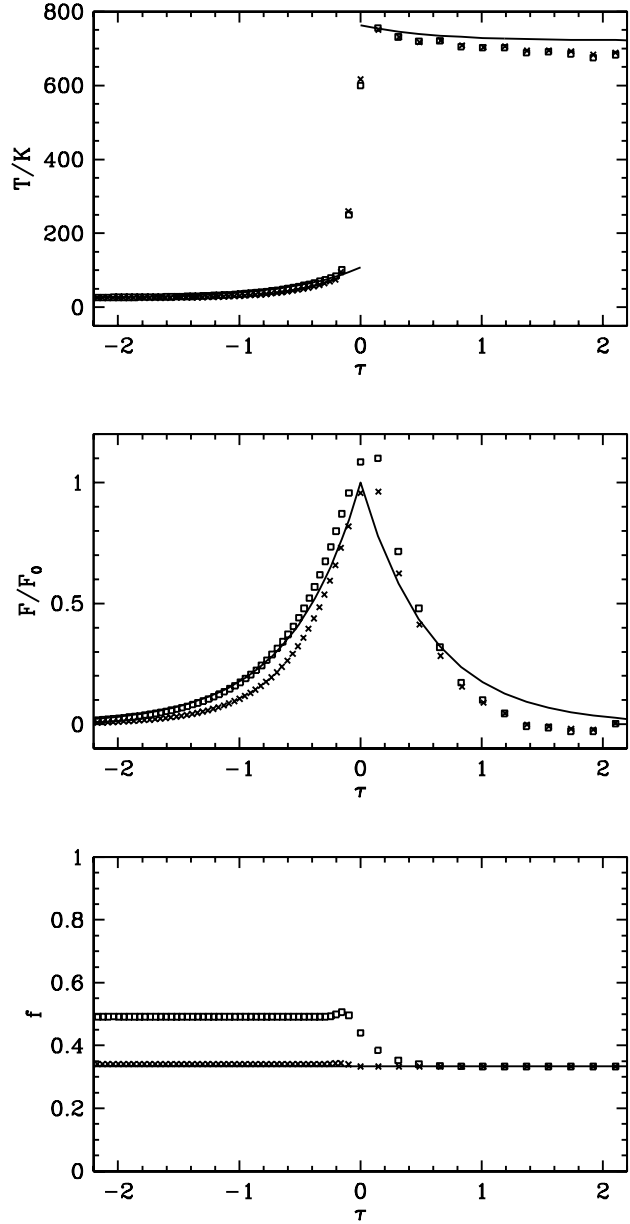}
\caption{\label{fig:subcritical}Temperature (top), flux (middle), and
Eddington factor (bottom) in a subcritical shock.  Optical depth
$\tau$ is measured from the shock front.  Parameters are the same as
for Sincell et al. (1999) figures 4 and 8.  Results obtained using the
LP flux limiter are shown by squares, Minerbo limiter by crosses.
Approximate analytic solutions from Zel'dovich \& Raizer (1967) are
indicated by solid lines.  Flux is normalized to the value at the
front in the analytic solutions.  Solid line in the bottom panel
indicates Eddington factor $\frac{1}{3}$.  }
\end{figure}

Results for a subcritical shock, with piston speed 6~km~s$^{-1}$, are
shown in figure~\ref{fig:subcritical}.  The time is $10^4$ seconds.
Temperature declines from the post-shock maximum more rapidly than
expected, and the flux is too large in several zones downstream from
the front.  These effects may be due to an incorrect angular
distribution of specific intensity.  They are similar when the
numerical resolution is doubled, suggesting they are not due to the
finite optical depth of the front imposed by the artificial
hydrodynamic viscosity.  Sincell et al. (1999) found, by directly
solving the transfer equation, that the Eddington factor near the
shock front is slightly less than $\frac{1}{3}$ because the emitting
layer has greatest line-of-sight optical depth when seen at grazing
angles.  In the FLD approximation, angular variation is assumed to be
either smooth (LP, eq.~[\ref{eqn:lpangular}]) or stepwise
(Minerbo), and this feature is missed.  Upstream, flux is higher with
the LP than with the Minerbo flux limiter.  As may be seen from
figure~\ref{fig:limiters}, the flux is reduced below the Eddington
value at larger optical depths with the Minerbo limiter.

\begin{figure}
\plotone{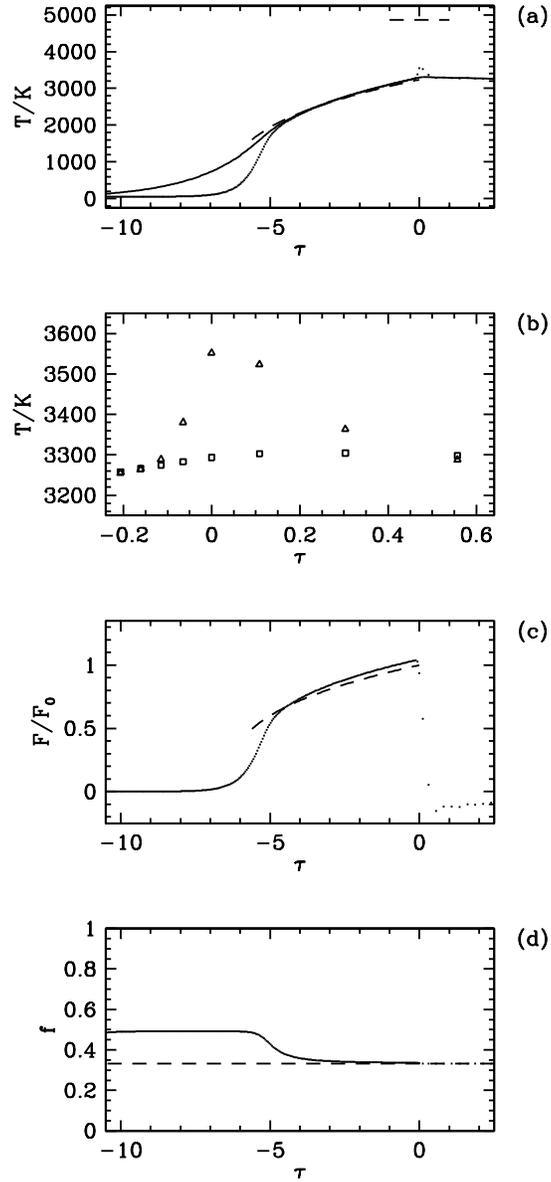}
\caption{\label{fig:supercritical}Gas and radiation temperatures (a),
closeup of the post-shock temperature spike (b), flux (c), and
Eddington factor (d) versus optical depth in a supercritical shock,
with LP flux limiter.  These may be compared with Sincell et
al. (1999) figures 7 and 9.  Radiation temperature in (a) is shown by
a solid line, gas temperature by points.  An approximate analytic
solution from Zel'dovich \& Raizer (1967) is indicated by a dashed
curve, and horizontal dashed line indicates the predicted height for
the temperature spike.  Radiation temperatures in (b) are shown by
squares, gas temperatures by triangles.  Flux in (c) is normalized to
the value at the front in the analytic solution.  Dashed line in (d)
indicates Eddington factor $\frac{1}{3}$.  }
\end{figure}

The structure calculated for a supercritical shock with piston speed
16~km~s$^{-1}$ is shown in figure~\ref{fig:supercritical}.  The time
is again $10^4$ seconds.  The temperature spike at $\tau=0$ is about
one-fifth as tall as found by Sincell et al. (1999), and about five
times as wide, due to the limited spatial resolution.  The peak
temperature in the spike is 314~K above the downstream temperature
$T_1$, whereas Zel'dovich \& Raizer estimate that the spike amplitude
is $(\frac{4}{\gamma+1}-1)T_1 = 1619$~K.  With spatial resolution
doubled, the spike is narrower and its height is larger at 506~K.  The
feature has approximately the expected degree of asymmetry, extending
to a larger optical depth downstream than upstream.  The flux in the
equilibrium preheated region is slightly overestimated, but away from
the front varies with optical depth in a manner close to that
predicted by Zel'dovich \& Raizer (1967).  The flux upstream from the
equilibrium region depends on the choice of limiter, with the Minerbo
limiter producing smaller flux and lower Eddington factor, as in the
subcritical case.

In both subcritical and supercritical shock calculations, the outcome
is independent of the timestep for steps smaller than about ten zone
radiation diffusion times $(\Delta x)^2\over D$.  We conclude from
these results that on a uniformly spaced grid, the FLD approximation
is a poor choice for determining detailed structure of shocks of
intermediate optical depth.  It may be sufficiently accurate for
multi-dimensional calculations in which such shocks must be followed
approximately as part of a more complex flow.

\subsection{Radiation-Dominated Shock \label{sec:raddomshock}}

\begin{figure}
\plotone{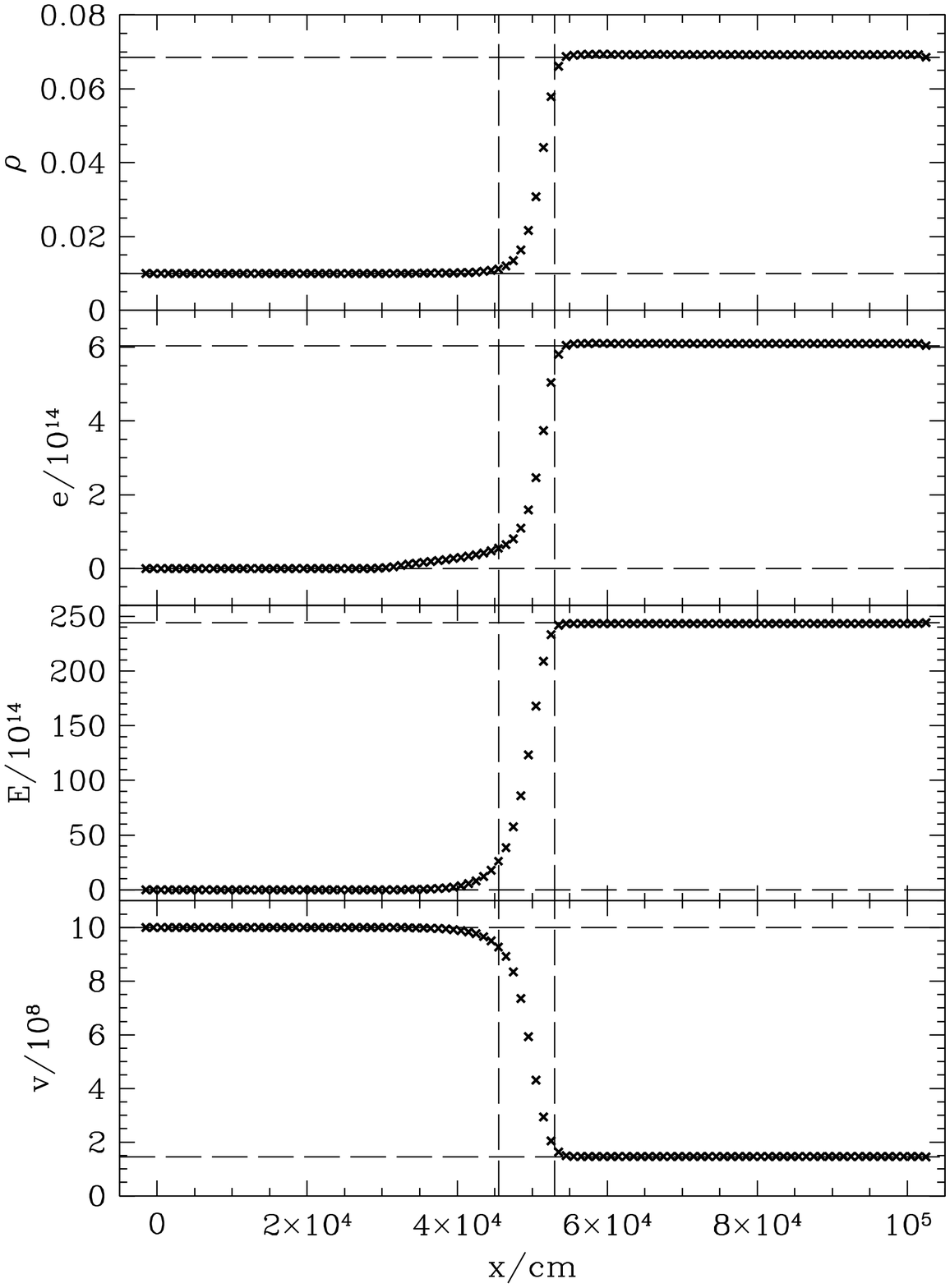}
\caption{\label{fig:raddomshock}Shock of Mach number 658 in an
optically-thick absorbing medium.  Horizontal dashed lines show
upstream and downstream states expected from the jump conditions.
Vertical dashed lines indicate the approximate expected extent of the
preheated layer ahead of the shock.  Optical depth between the
vertical lines is about 80.  Each cross marks one of the 100 grid
zones.  Quantities plotted, from top to bottom, are the density,
thermal and radiation energy densities, and velocity, in cgs units.  }
\end{figure}

If material upstream and downstream from a radiating shock have
sufficient optical depth, radiation is trapped near the interface,
diffusing a limited distance upstream.  In this section we examine
whether the code adequately matches the jump conditions and thickness
of such a shock.  Parameters are chosen so the energy density is
mostly in the gas upstream and in the radiation downstream.  Opacity
is assumed to be due to absorption as in \S\ref{sec:critical}, and
material and radiation are assumed to be in thermal equilibrium far
from the shock.  Initially, the left of the grid is set to density
0.01~g~cm$^{-3}$, temperature $10^4$~K, and speed $10^9$~cm~s$^{-1}$,
yielding a Mach number of 658.  The right is set to density
0.0685847~g~cm$^{-3}$, temperature $4.239\times 10^7$~K, and speed
$1.458\times 10^8$~cm~s$^{-1}$, as computed from the jump conditions.
The boundary conditions are inflow on the left, and outflow on the
right.  After a brief transient, a steady shock is established.  The
situation after 50 flow crossing times is shown in
figure~\ref{fig:raddomshock}.  The approximate expected thickness of
the shock is the distance $l$ for which the time to diffuse upstream,
$l^2/D$, matches the time to sweep downstream, $l/u_1$ (Mihalas \&
Mihalas 1984).  Here $D$ is the radiation diffusion coefficient and
$u_1$ the upstream gas speed.  The shock thickness obtained in the
calculation is consistent with the resulting estimate of 7500~cm.  The
values downstream in the calculation differ by a maximum of 1.0\% from
those predicted using the jump conditions.  When zone spacing is
decreased a factor four, the discrepancies are reduced to 0.2\%,
indicating that mass, momentum, and energy are adequately conserved in
material passing through the shock.

\section{SUMMARY\label{sec:summary}}

We have described a set of subroutines which may be used with the
ZEUS-2D magnetohydrodynamics code for radiation MHD calculations in
the flux-limited diffusion approximation.  Results were obtained
typically in three to ten times more floating-point operations per
timestep than required for similar calculations without radiation,
with the greater number needed in the optically-thin tests.  The
primary limitation of flux-limited diffusion is that the flux is not
evolved as an independent variable, but is assumed anti-parallel to
the gradient in radiation energy density.  This is incorrect in flows
where shadows are cast or inclined radiation fronts cross.  The
incorporation in the flux limiter of a specified angular dependence
for the radiation field has greatest effect in regions of intermediate
to low optical depth, and may contribute to the broadening of
radiation fronts.  Flux limiters derived for angular dependences
occurring in particular applications may be useful in obtaining more
accurate solutions.

The limitations of FLD revealed in the tests mean that care must be
taken in applying the module to problems with highly anisotropic
radiation fields in optically thin regions.  In such cases, full
transport methods may be more appropriate.  However, in optically
thick flows, FLD is more efficient.  The module has proven useful in
studies of the internal dynamics of optically thick,
radiation-dominated accretion disks (Agol et al. 2001; Turner \& Stone
2001).  Although the method remains stable for long timesteps, it
shares with other implicit schemes the property that fluctuations with
periods less than the timestep may be rapidly and unphysically damped,
as seen in the linear wave and sub- and supercritical shock tests.
When the diffusion timestep constraint of \S\ref{sec:dt} is violated,
it is advisable to check that results are independent of the timestep.

The radiation terms included in the equations solved are those of
order unity in $v/c$ in at least one of the free-streaming, static
diffusion, and dynamic diffusion regimes (Mihalas \& Mihalas 1984;
Stone, Mihalas, \& Norman 1992).  Adding higher-order terms would
allow study of special relativistic effects such as radiation drag and
photon viscosity.  In the implementation described here, the radiation
source terms are operator-split, with the flux divergence term evolved
separately from the remainder.  Linear waves and shocks to which both
groups of terms contribute were evolved adequately, but situations may
be encountered where the operator splitting leads to inaccuracies.  We
have assumed the gas is in local thermodynamic equilibrium, and
relaxing this assumption may in the future allow a great variety of
additional problems to be addressed.

\begin{acknowledgments}
Financial support of this work from the United States Department of
Energy is gratefully acknowledged.  We thank Eric Agol and Julian
Krolik for their contributions to the tests of the module, and the
anonymous referee for helpful comments on the presentation and the
importance of the matrix condition numbers.
\end{acknowledgments}

\end{document}